\documentclass[sigconf, nonacm, timestamp]{acmart}

\AtBeginDocument{%
  \providecommand\BibTeX{{%
      \normalfont B\kern-0.5em{\scshape i\kern-0.25em b}\kern-0.8em\TeX}}}

\usepackage{graphicx}
\usepackage{textcomp}
\usepackage{multirow}
\usepackage{color}
\usepackage[ruled, vlined]{algorithm2e}
\usepackage{epstopdf}
\usepackage{comment}
\usepackage{multirow}
\usepackage{url}
\usepackage{subcaption}
\usepackage{tabulary}
\usepackage[newcommands]{ragged2e}
\usepackage[export]{adjustbox}

\newcommand{\our}{\emph{SenseRT}\xspace}

\newcommand{\one}{({\em i})\/}
\newcommand{\two}{({\em ii})\/}
\newcommand{\three}{({\em iii})\/}
\newcommand{\four}{({\em iv})\/}
\newcommand{\five}{({\em v})\/}

\newcommand\s[1]{(\S\ref{#1})\xspace}

\setcopyright{acmcopyright}
\copyrightyear{2018}
\acmYear{2018}
\acmDOI{10.1145/1122445.1122456}

\acmConference[Woodstock '18]{Woodstock '18: ACM Symposium on Neural
  Gaze Detection}{June 03--05, 2018}{Woodstock, NY}
\acmBooktitle{Woodstock '18: ACM Symposium on Neural Gaze Detection,
  June 03--05, 2018, Woodstock, NY}
\acmPrice{15.00}
\acmISBN{978-1-4503-9999-9/18/06}

\begin{document}

\title{\our: A Streaming Architecture for Smart Building Sensors}

\author{Rohit Verma}
\email{rv355@cam.ac.uk}
\affiliation{%
  \institution{University of Cambridge}
  \city{Cambridge}
  \country{United Kingdom}
}

\author{Justas Brazauskas}
\email{jb2328@cam.ac.uk}
\affiliation{%
  \institution{University of Cambridge}
  \city{Cambridge}
  \country{United Kingdom}
}

\author{Vadim Safronov}
\email{vs451@cam.ac.uk}
\affiliation{%
  \institution{University of Cambridge}
  \city{Cambridge}
  \country{United Kingdom}
}
\author{Matthew Danish}
\email{mrd45@cam.ac.uk}
\affiliation{%
  \institution{University of Cambridge}
  \city{Cambridge}
  \country{United Kingdom}
}
\author{Jorge Merino}
\email{jm2210@cam.ac.uk}
\affiliation{%
  \institution{University of Cambridge}
  \city{Cambridge}
  \country{United Kingdom}
}
\author{Xiang Xie}
\email{xx809@cam.ac.uk}
\affiliation{%
  \institution{University of Cambridge}
  \city{Cambridge}
  \country{United Kingdom}
}
\author{Ian Lewis}
\email{ijl20@cam.ac.uk}
\affiliation{%
  \institution{University of Cambridge}
  \city{Cambridge}
  \country{United Kingdom}
}
\author{Richard Mortier}
\email{rmm1002@cam.ac.uk}
\affiliation{%
  \institution{University of Cambridge}
  \city{Cambridge}
  \country{United Kingdom}
}


\begin{abstract}
  Building Management Systems (BMSs) have evolved in recent years, in ways that require changes to existing network architectures that follow the store-then-analyse approach. The primary cause is the increasing deployment of a diverse range of cost-effective sensors and actuators in smart buildings that generate real-time streaming data. Any in-building system with a large number of sensors needs a framework for real-time data collection and concurrent stream processing from sensors connected using a range of networks.

  We present \our, a system for managing and analysing in-building real-time streams of sensor data. \our collects streams of real-time data from sensors connected using a range of network protocols. It supports concurrent modules simultaneously performing stream processing over real-time data, asynchronously and non-blocking, with results made available with minimal latency. We describe a prototype implementation deployed in two University department buildings, demonstrating its effectiveness.
\end{abstract}

\ccsdesc[300]{Networks~Sensor networks}
\ccsdesc[500]{Hardware~Sensors and actuators}
\ccsdesc[500]{Computer systems organization~Real-time system architecture}

\keywords{real-time data, stream processing, smart buildings, sensors, spatio-temporal data}

\maketitle

\section{Introduction}

Increased penetration of the Internet of Things (IoT) in our lives~\cite{dahlqvist2019growing}, is making it increasingly easy to deploy energy-efficient sensors and actuators to perform tasks like occupancy detection~\cite{dodier2006building} or indoor-environment control~\cite{cao2009building}. The number of deployed IoT devices by 2021 just for smart buildings would be around $10.8$ billion ($2.8$ billion for residential buildings)~\cite{memoori2018}. To put that in perspective, an average home in 2020 would generate approximately $4.7$ terabytes of data annually~\cite{memoori2018,audienceproject}. Handling such amounts of data is a challenging problem, exacerbated by most such sensors generating real-time streams of data.

\begin{table*}
  \begin{tabulary}{\linewidth}{L|L|L|L|L|L}
    \textbf{Protocol}  & \textbf{Frequency}     & \textbf{Data Rate}     & \textbf{Range} & \textbf{Power} & \textbf{Security}          \\ \hline
    \textbf{Bluetooth~\cite{raza2015bluetooth}} & 2.4 GHz                & 1 Mb/s                 & 50 m           & Low            & 64 and 128 bit encryption  \\ \hline
    \textbf{Wi-Fi~\cite{alliance2009wi}}     & 2.5 GHz, 5 GHz          & 150-200 Mb/s (typical) & 50--100 m      & High           & 256 bit key encryption     \\ \hline
    \textbf{ZigBee~\cite{alliance2010zigbee}}    & 2.4 GHz                & 250 kb/s               & 10--100 m     & Low            & 128 AES layer security     \\ \hline
    \textbf{LoRa~\cite{bor2016lora}}      & Region specific & 0.3--50 kb/s          & 2--5 km       & Low            & 128 bit AES encryption key \\ \hline
    \textbf{Modbus~\cite{modbus}}    & 2.4 GHz & 9600 b/s          &  Layout dependent       & High            & 128 bit AES encryption (if Modbus/TLS) \\
    \hline
  \end{tabulary}
  \caption{\label{tab:comprot}Popular sensor communication protocols used in smart buildings}
\end{table*}

The data generated has both spatial and temporal aspects. For example, the humidity level falling below a threshold causing discomfort to occupants will be observed by a sensor deployed in a specific place in the building (spatial aspect) at a particular date and time (temporal aspect). The spatial aspect is relatively straightforward to handle, but there are several subtleties to the temporal aspect: what time should be assigned as the \emph{time of reading}, and associated with a particular sensor reading? When the data was sensed, when the sensor transmitted the reading, or when it arrived at the receiving platform?

\begin{figure}[!ht]
  \centering
  \includegraphics[width=\linewidth]{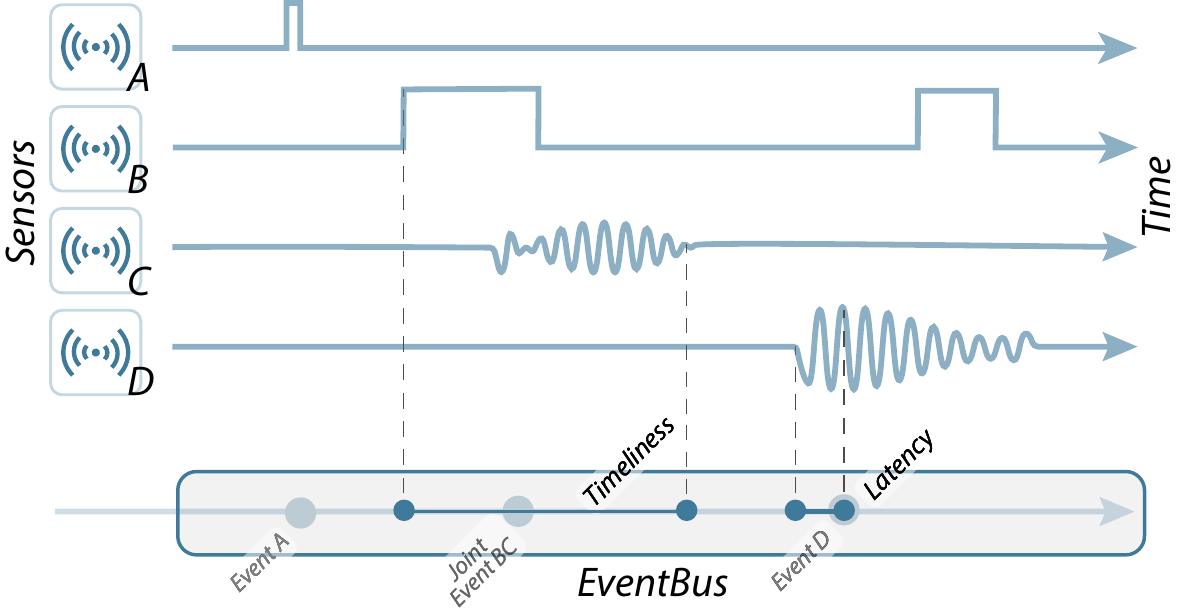}
  \caption{\label{timeliness}Timeliness vs latency.}
  \vspace{-3ex}
\end{figure}

This brings to the fore concepts of \emph{latency} and \emph{timeliness}, depicted in Figure~\ref{timeliness}, often incorrectly considered synonymous. We use \emph{latency} to refer to the delay introduced between the time of recording the reading at the sensor and that reading arriving at the intended destination. In contrast, \emph{timeliness} refers to the inherent characteristic of an event, the timescale on which it is appropriate to determine a reading changed. In the humidity example above, \emph{latency} would be the delay between the time the sensor recorded the humidity and when the system determining the degree of occupant comfort received the reading; this will vary between sensors and over time. However, \emph{timeliness} is the time period over which readings from the sensor lead to a situation where the humidity threshold overshoots. A reading from the sensor that arrives after this time period won't help in computing the event of discomfort. Any framework striving for efficient real-time data management and analysis needs to treat these concepts with care.

Furthermore, the concept of \emph{timeliness} also explains the need of moving towards a stream processing based approach from the prevalent store-then-analyse approach~\cite{ortiz2013platform,dawson2010smap}. These existing approaches fail to provide real-time insights when the data never stops and usefulness of the streaming data is ephemeral but has to adhere to \textit{timeliness} requirements~\cite{swim}. Nowadays sensors are used to manage and control several crucial aspects of a building like identifying time-critical events of gas leaks~\cite{garabedian2006automated} or fires~\cite{luo2002fire}. In such scenarios, any delay between generation of data at the sensor and its analysis should be minimised as much as possible, wherein stream processing~\cite{stonebraker20058} comes in.

Building a system to handle devices in a building and the large volume of real-time data streams they generate while keeping these key spatio-temporal concepts in mind has multiple challenges. First, the heterogeneous nature of sensors and consequently the data they generate. The deployed sensors in a building could connect using channels including Bluetooth, Wi-Fi, ZigBee, LoRaWAN, or the in-building Modbus. Moreover, the definition of time of reading for sensors using these different networks could be different. The system should accept data from any such channels and normalise aspects such as time of reading. Second, analysis of the stream of data generated from all the in-building sensors in real-time should introduce minimal latency between data generation, data analysis, and publication of a result, especially for crucial events such as fire or power outage in critical areas. Moreover, processing of this stream of real-time data should be done for all sensors concurrently. Third, all data flow must be asynchronous and non-blocking while supporting concurrency.

To tackle these challenges, we develop \our, a system for real-time data flow in smart buildings. Key contributions include:
\begin{itemize}
\item \our defines a set of design principles  ensuring that the architecture is easy to mutate based on the needs of the building at any time of building design.
\item \our provides custom decoders to retrieve data from sensors on different channels, accumulated via bridging framework of Message Queuing Telemetry Transport (MQTT)~\cite{standard2014mqtt}, and normalise it for easy use throughout the system.
\item \our provides a suite of modules which follow the \textit{stream processing}~\cite{stonebraker20058} framework to ensure that the streams of real-time data from a large number of sensors is managed and analysed with minimal latency and adhere to timeliness bounds.
\item All data handling and stream processing modules are designed to follow the \textit{actor model}~\cite{hewitt2010actor}, which guarantees asynchronous and concurrent processing. We modify the actor model such that the concurrent modules could work in a non-blocking fashion.
\item \our provides support for client-side applications to analyse and process the data as they need.
\end{itemize}

We continue by discussing related work in smart buildings~\s{s:relatedwork}, before describing the design principles and high-level system architecture~\s{s:architecture}. We implemented a prototype of \our in two buildings in our university and ran experiments to observe the architecture's effectiveness for seven months. We describe this implementation~\s{s:implementation} and a case study of an end-to-end application~\s{s:casestudy}, before evaluating the implemented prototype~\s{s:evaluation}. We conclude with a discussion of limitations and future work~\s{s:conclusion}.

\section{Related Work}\label{s:relatedwork}

The IoT market is expected to increase to 5.8 billion endpoints in 2020, a $21\%$ increase from 2019, with utilities (electricity, water) the most significant users (1.17 billion) and building automation showing the highest growth of $42\%$~\cite{gartner}. A boost towards this direction could be because of the growth in the number of smart devices and smart objects that build the ecosystem for a smart building~\cite{dave2011next}. This ecosystem of smart devices and smart objects is enabled by the network technology use, with different protocols (e.g.,~LoRaWAN, ZigBee, Wi-Fi, Bluetooth) being selected based on the desired goals.

\subsection{Communication in Smart Buildings}
Communication protocols enable the exchange of a massive stream of data between sensors and the network.  Factors like range, data load, power demand, and security define which communication protocol would be suitable for a particular set of smart devices. A comparison of the major protocols is given in Table~\ref{tab:comprot}.

There has been quite a focus on using LoRa for smart buildings~\cite{li2019indoor, havard2018smart} due to its low power consumption, low cost, long-range, bi-directional communication made possible by chirp spread spectrum (CSS)~\cite{nguyen2019efficient}, standardization and flexibility of selecting bandwidth, code rate and spreading factor as per need. Unlike the other protocols, Modbus is a wired protocol for industrial automation systems that has recently become popular for building management systems, especially for meters and HVACs.

\subsection{Sensing in Buildings}
Building Energy and Comfort Management (BECM)~\cite{shaikh2014review} has been a significant cause of the introduction of sensing in buildings. BECM tries to achieve optimal energy consumption~\cite{spataru2014monitor, zhang2019cooperative} and provide a high level of indoor environment quality (IEC)~\cite{al2016occupant} by using different types of sensor. We divide them into two broad categories:

{\bf Dumb Sensors}. This type includes sensors such as pressure mats, IR sensors, $CO_2$ sensors, temperature sensors, particulate matter sensors. They  simply read and transmit readings for values they sense, and can be used to understand occupant behaviour patterns~\cite{gao2009self, zhang2019domain} or to optimise IEC parameters to improve thermal comfort, visual comfort, or indoor air quality~\cite{dong2009sensor, nassif2012robust}.

{\bf Intelligent Sensors}. This type compromise one or more \textit{dumb sensors} with attached computation that can process the raw data and relay analysed results. Example include smartphones, fingerprint sensors, wearable sensors, smart cameras, and body thermometers. In smart buildings these sensors can be used for traditional purposes such as occupancy detection and managing IEC~\cite{ye2003new}, but can also provide more personalised functions~\cite{sim2016estimation}.

\subsection{Smart Building Management and Control}
Several works provide a framework for building control to balance energy consumption and maintain occupant comfort. iDorm~\cite{hagras2004creating} set up a testbed where multiple embedded sensors were fitted in a dorm room to obtain responsive inputs from the user, which helped to learn user preferences using distributed AI and fuzzy-genetic logic. MASBO~\cite{liu2008multi} is a multi-agent system where data arriving from a Building Management System (BMS) is observed and analysed by a set of agents to provide suitable energy-efficient control for the building without compromising on occupant comfort. Chen et al.~\cite{chen2009design} propose a hierarchical system architecture that emphasises improving savings over the building life-cycle while addressing stakeholder goals.

Another class of work concentrates more on how sensor data could be collected and managed in a smart building. Choubey et al.~\cite{choubey2015power} set up a localised sensor network in an area and perform localised data processing for this set of sensors. LabVIEW~\cite{shah2016customized} provides a data collection framework to collect humidity, temperature, and light data from sensors in a wireless sensor network in the building. Bashir et al.~\cite{bashir2016towards} provide an IoT Big Data Analytics (IBDA) based framework for storage and analysis of real-time data that the IoT sensors in a smart building generate.

Most architectures follow the trend of using a particular approach to collect sensor data locally and then store it on the cloud for further processing. Data collection efficiency is achieved by using different means like adding a programmed data acquisition chip~\cite{al2017smart}, setting up a fog server~\cite{fayyaz2019iot}, or using IPv6 over Low power Wireless Personal Area Networks (6LoWPAN)~\cite{evangelatos2012evaluating}. However, all of these systems rely on stored data to perform any analysis rather than analysing it as and when it arrives.

\subsection{Necessity of Stream Processing}

\begin{figure}[!ht]
    \includegraphics[width=\linewidth]{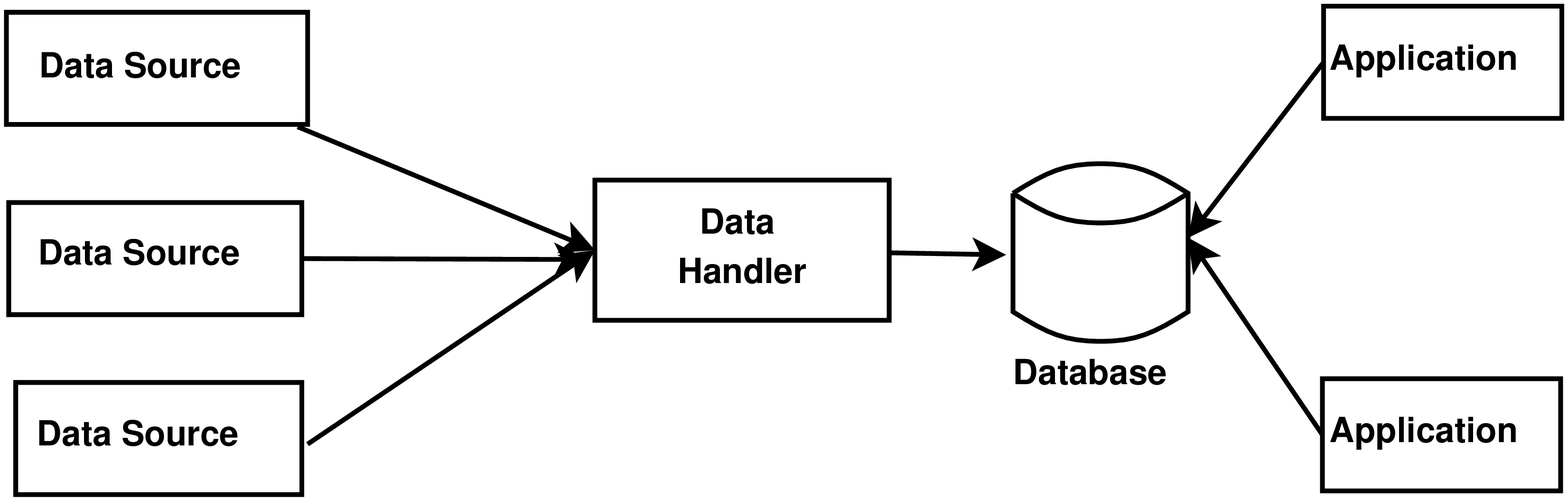}
    \caption{\label{f:sta}Store-then-Analyze Model}
\end{figure}

The existing systems~\cite{shah2016customized, bashir2016towards, al2017smart, fayyaz2019iot, dawson2010smap}, utilize the prevalent store-then-analyze architecture, where data is first stored in a data repository and then analyzed as per the needs of the applications (Figure~\ref{f:sta}). However, several use-cases of sensing in buildings are linked to time-critical responses. It is very important to minimize delay in safety based use-cases like gas leak~\cite{garabedian2006automated}, water leak~\cite{mccarthy2018detection} or fire detection~\cite{luo2002fire} in buildings. When considered as basic components of a smart-grid network, timely analysis of energy consumption is important to optimize smart-grid management~\cite{costanzo2012system,morvaj2011demonstrating}. When the building is used for specialized purposes like hospitals~\cite{ko2010medisn} or elderly care homes~\cite{suryadevara2012wireless, alam2019besi}, adhering to timeliness of sensor data becomes crucial. When trying to act on such scenarios, which generate always moving time-critical data, the store-then-analyze approaches fall short. The reliance on a source of data repository would involve several to-and-fro network transactions resulting in crucial time loss. This fails a real-time system's adherence to timeliness~\cite{swim, stonebraker20058}.

Along the line of these time-critical solutions, ScaleOut~\cite{scaleout} has pointed out the importance of real-time stream processing~\cite{stonebraker20058} for digital twins. ThoughtWire~\cite{thoughtwire} who also work with stream processing based systems for digital twins share a similar notion. CityPulse~\cite{tonjes2014real} and Zhou et al.~\cite{zhou2016iot} advocate the necessity of stream processing based data analytics for smart city projects.

\begin{figure}[!ht]
    \includegraphics[width=\linewidth]{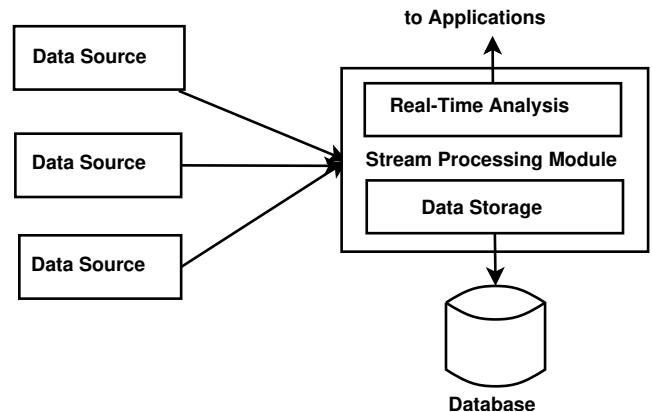}
    \caption{\label{f:sp}Stream Processing}
\end{figure}

Figure~\ref{f:sp} shows the stream processing approach which unlike Figure~\ref{f:sta} performs storage and analysis simultaneously. The stream processing module has two (or more) concurrent processors which analyse and store data at the same time. The analysed results are then made available to any applications which have subscribed a priori.

\section{Design \& Architecture}\label{s:architecture}

We first describe the design principles \our is built upon and then its high-level architecture.

\subsection{Design Principles}
The \our architecture must be flexible to changes at any period of deployment and robust enough to handle the continuous stream of real-time data entering the system through the numerous sensors in the building via different channels. To achieve this, we apply the following design principles.

{\bf All data is spatio-temporal}. Over a long enough period, the metadata about any part of the building or sensors is bound to change. For instance, a sensor could be moved to a new room changing its location, or a room be divided into two rooms changing its boundary.  Tagging every datum with a timestamp and location information, enables change tracking of deployments.

{\bf Spatial hierarchy of container objects}. Sensors could be deployed in different locations having different usage and access control policies depending on the granularity of location considered (e.g.,~building, floor, room, desk). This defines each component as an object which could hold multiple other objects, each of which could have its own set of sensors. This ensures physical changes in the building are handled with minimal changes, e.g.,~changing only the parent when a desk is moved to another room, or reusing the floor model on adding a new floor to the building.

{\bf Stream processing of data}. \our uses a publish/subscribe model for data exchange, ensuring no polling is required for the real-time stream of data arriving at any time instance to the system. This reduces delivery latency, essential for real-time data analysis.

{\bf Asynchronous message transfer in a non-blocking framework}. Asynchronous message transfer ensures that any message from any sensor or system module could be analysed in real-time, and multiple modules can work on the same data concurrently. This in turn guarantees a non-blocking data processing framework.

\begin{figure}
  \centering
  \includegraphics[width=\linewidth]{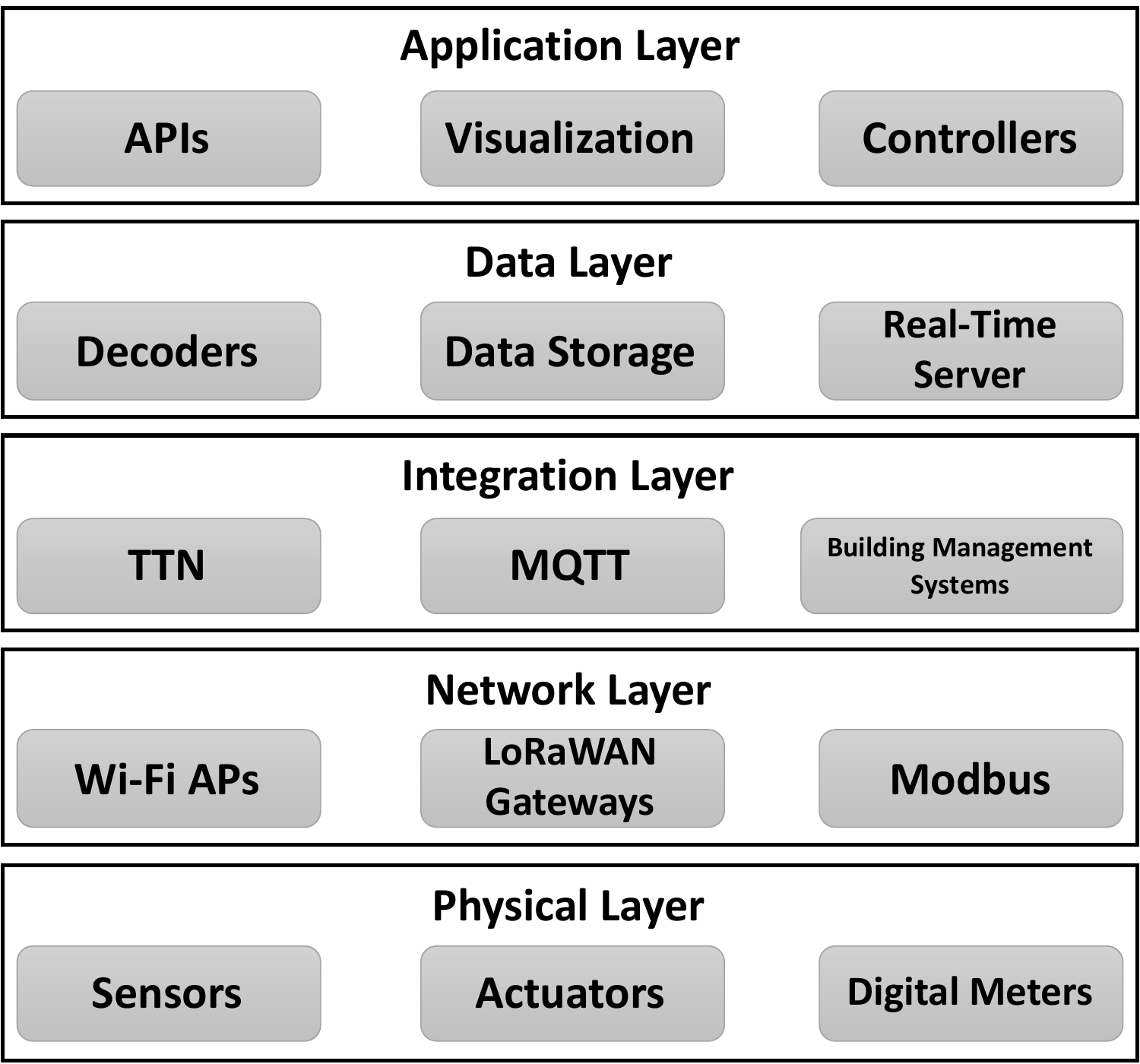}
  \caption{\label{arch}\our high level system architecture.}
  \vspace{-3ex}
\end{figure}

\subsection{System Architecture}
The high-level decomposition of \our results in five layers, shown in Figure~\ref{arch}: Physical, Network, Integration, Data, and Application.

{\bf Physical Layer}. Holds all hardware devices like sensors, actuators, and building meters. Devices report data and exchange messages using one or more of their supported protocols and formats. Our deployment uses a wide range of sensors outlined in Table~\ref{tab:sensorsused}.

{\bf Network Layer}. Houses the devices required to support message transfer from the sensors such as Wi-Fi access points, ZigBee or LoRaWAN gateways, or Modbus components. Our deployment uses Zigbee and Wi-Fi networks for short-range connectivity within particular areas, and a LoRaWAN network to provide backhaul interconnection within and between buildings.

{\bf Integration Layer}. Provides services to collate and homogenise data received over different network types from different hardware devices, making it much more straightforward to build applications to analyse and react to data from one or more sources. Consider a room that has smart plugs, LoRa sensors, and an electric meter. The smart plugs could be uploading data over Wi-Fi, LoRa sensors through LoRaWAN via The Things Network (TTN)~\cite{visser2018network}, and the electric meters through a Modbus. Services in this layer ensure that data received over all these protocols are available through a single channel. Furthermore, this layer also ensures that data exchange in the system is through a publish/subscribe method. Our deployment achieves this by making use of the Message Queuing Telemetry Transport (MQTT)~\cite{standard2014mqtt} and MQTT bridging.

{\bf Data Layer}. Provides services to manage data streams arriving from the integration layer. A set of decoders normalises the data from multiple devices, which is made available to a Real-Time Server (RTS). The RTS holds various crucial modules of the architecture. These modules handle real-time stream processing, routing messages to other similar systems as required, storing data for future usage, and making data available to the external application. In our deployment the RTS is implemented to be asynchronous and non-blocking using Vert.x. The data layer also houses the database, which stores the spatio-temporal metadata of all the sensors and the object-level components in a hierarchical structure.

{\bf Application Layer}. The client-facing layer, providing APIs and user interfaces for those wishing to access sensor data. Our deployment comprises a server that acts as the point-of-contact with \our for all client-side applications trying to access sensor data. The details of application architecture are out of scope for this paper as we are primarily concerned with the overall system.

\section{Implementation}\label{s:implementation}

\begin{figure*}
  \centering
  \includegraphics[width=\linewidth]{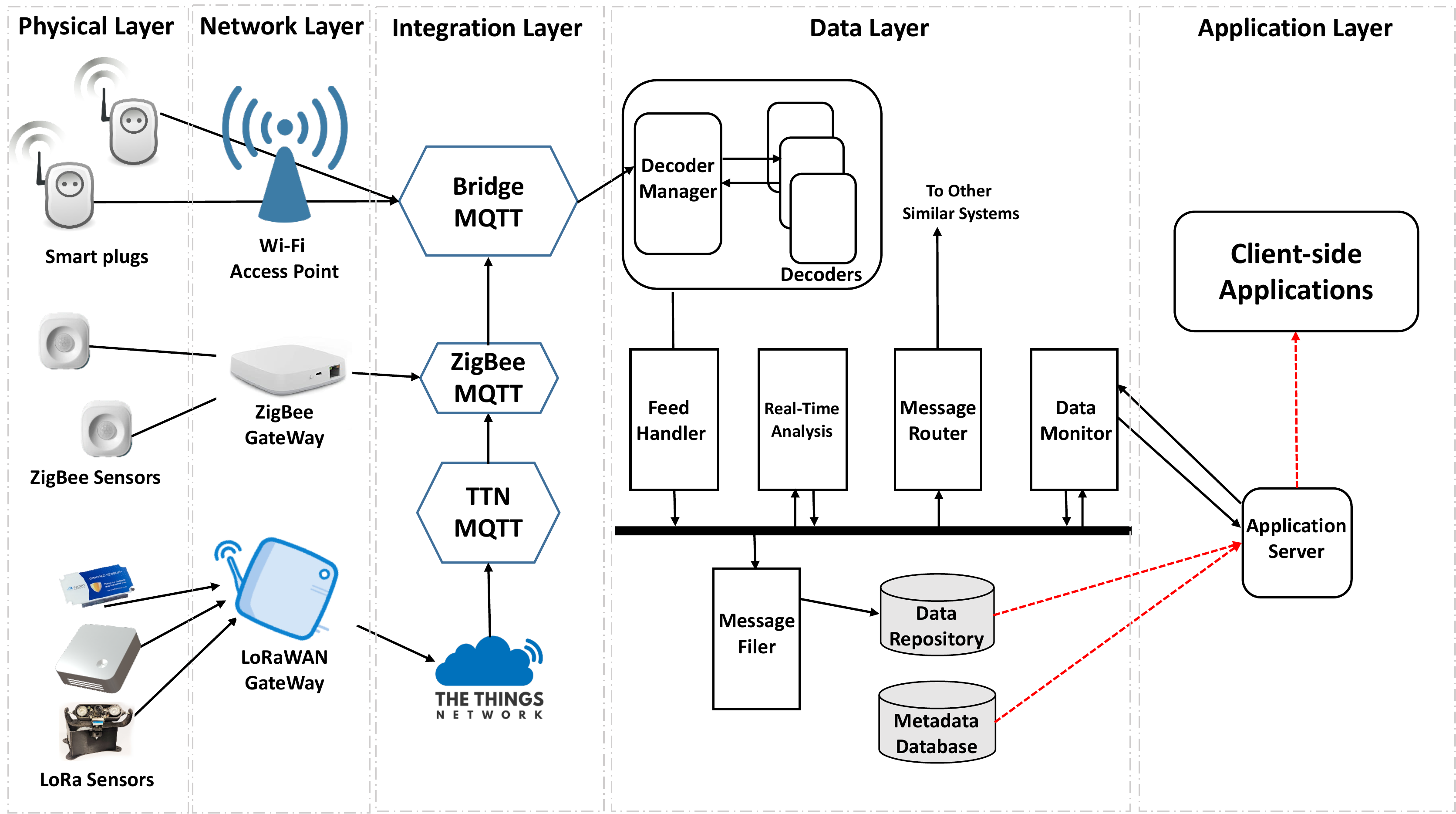}
  \caption{\label{f:implementation}Our prototype implementation of the \our architecture. Solid black lines show real-time data flow while dashed red lines show request-response data flow.}
  \vspace{-2ex}
\end{figure*}

We deployed a prototype of \our over two department buildings on our University campus. The two buildings each have multiple floors containing offices, labs, communal areas, and corridors. Figure~\ref{f:implementation} depicts the overall architecture of the prototype, and we detail each layer next.

\subsection{Physical Layer}

\begin{table}
  \begin{tabulary}{\linewidth}{p{6em}|p{10em}|p{4em}|r}
    \textbf{Sensor} & \textbf{Measures} & \textbf{Channel} & \textbf{Cost} \\ \hline
    Smart plugs & power & Wi-Fi & \$45 \\ \hline
    Infrared Motion & motion & ZigBee & \$17 \\ \hline
    Door/Window & open/close events & ZigBee & \$9\\ \hline
    CO$_2$ & CO$_2$, humidity, temperature, light, motion & LoRaWAN & \$205 \\ \hline
    Temperature & temperature & LoRaWAN & \$129 \\ \hline
    Tilt & tilt angle & LoRaWAN & \$99 \\ \hline
    Door/Window & open/close events & LoRaWAN & \$ 99 \\ \hline
    Water Leak & presence of water & LoRaWAN & \$119 \\ \hline
    Occupancy & occupancy, temperature, humidity, light, motion & LoRaWAN & \$123 \\ \hline
    DeepDish & people count & Wi-Fi & \$100 \\
    \hline
  \end{tabulary}
  \caption{\label{tab:sensorsused}Sensors used in our deployment.}
  \vspace{-2ex}
\end{table}

Table~\ref{tab:sensorsused} lists the sensors used in our deployment, dumb and intelligent.

\paragraph{Dumb Sensors}

\begin{figure}
  \centering

  \begin{subfigure}[t]{.24\linewidth}
    \includegraphics[width=\linewidth]{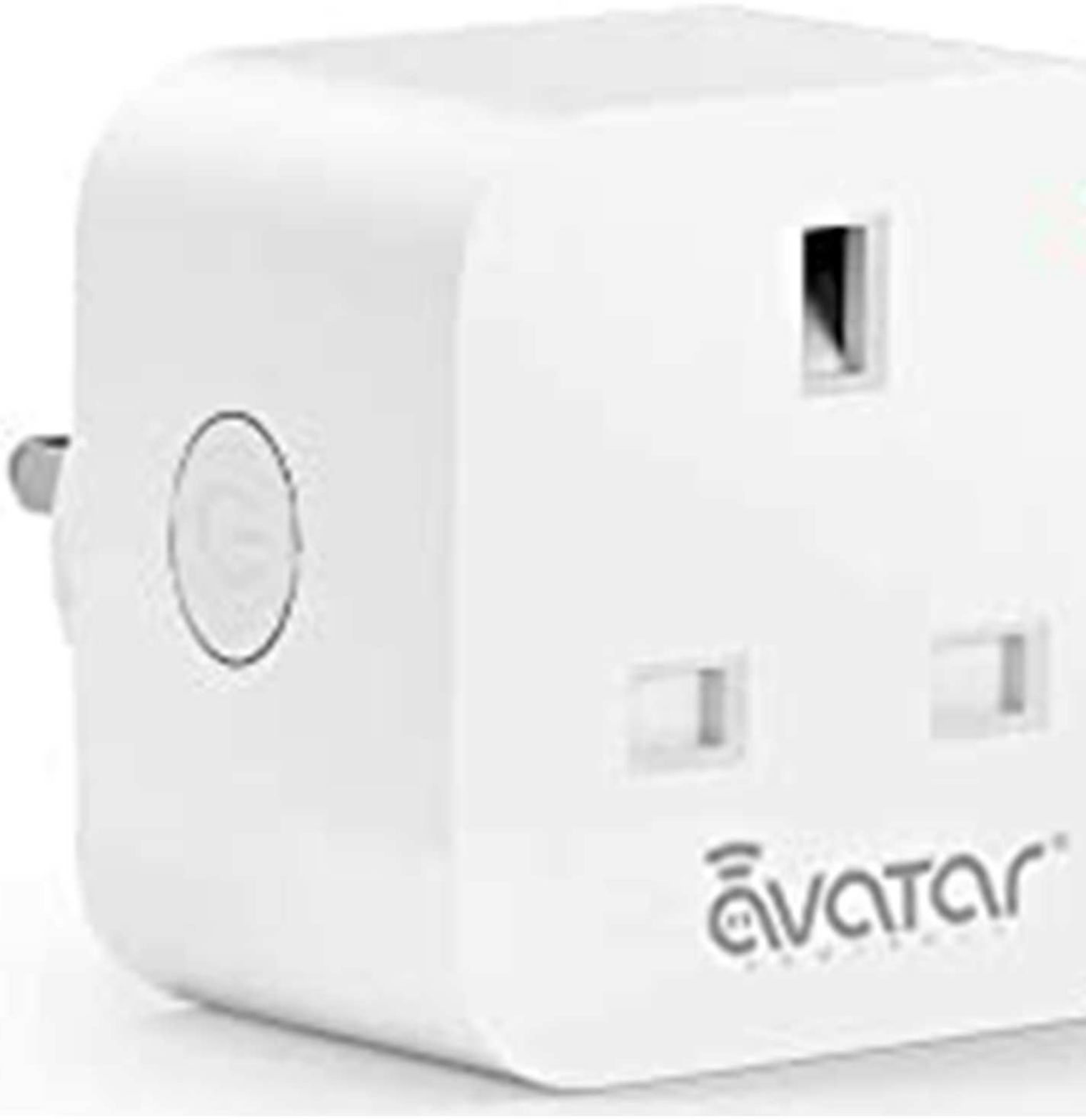}
    \caption{\label{f:smartplug}Smart plug}
  \end{subfigure}
  \hfill
  \begin{subfigure}[t]{.24\linewidth}
    \includegraphics[width=\linewidth]{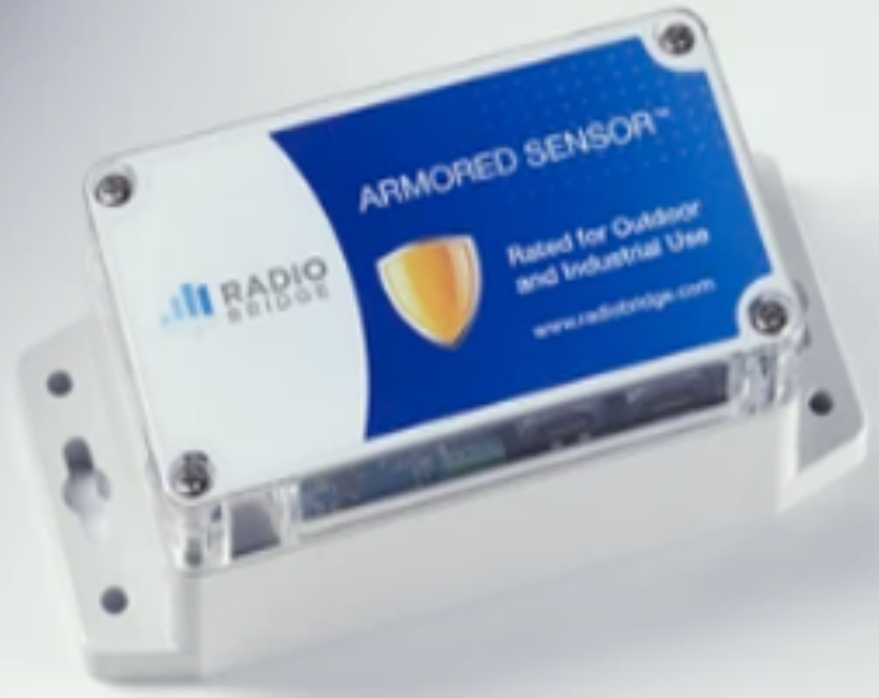}
    \caption{\label{f:lorasensor}LoRa}
  \end{subfigure}
  \hfill
  \begin{subfigure}[t]{.24\linewidth}
    \includegraphics[width=\linewidth]{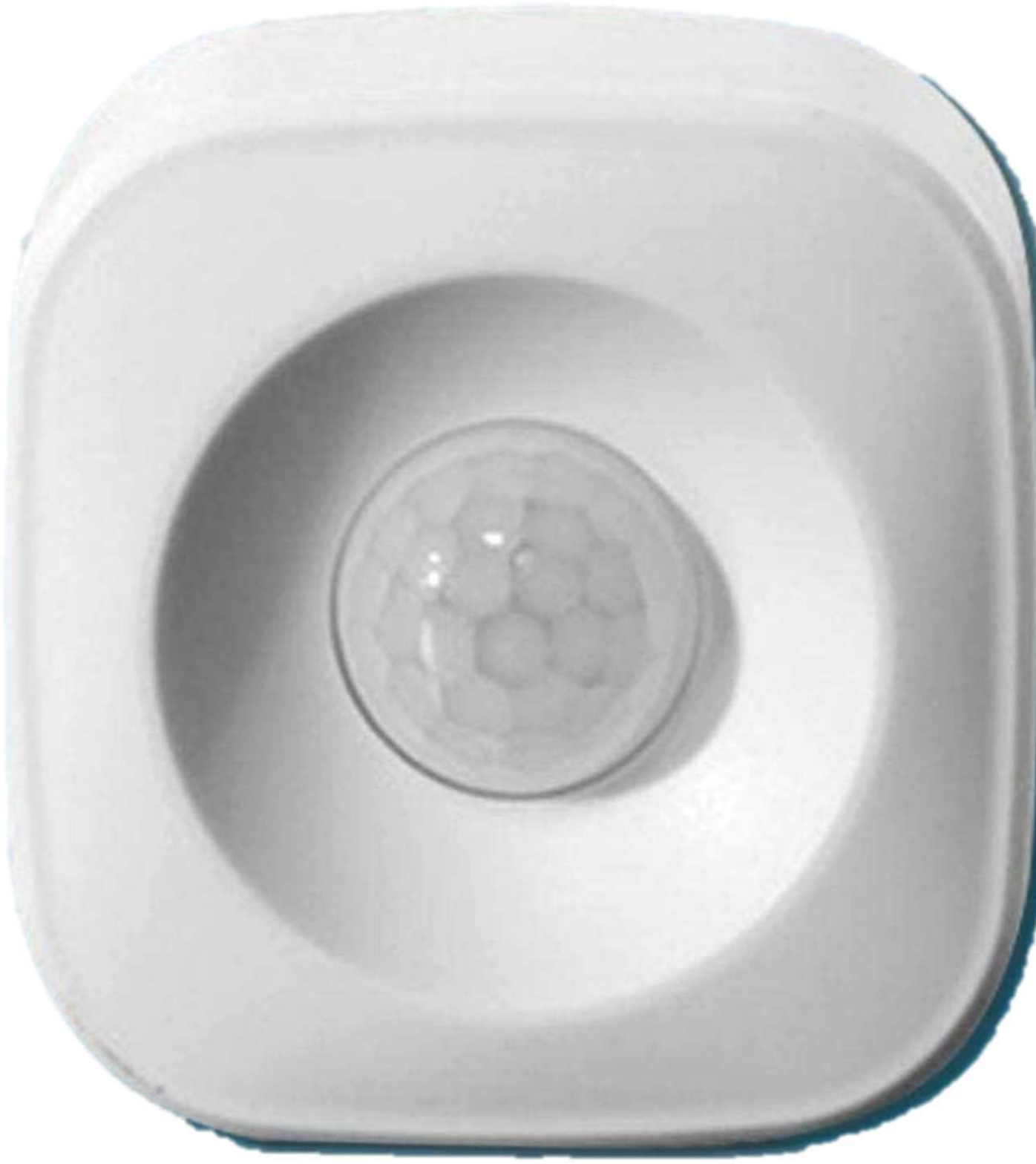}
    \caption{\label{f:zigbeesensor}ZigBee}
  \end{subfigure}
  \hfill
  \begin{subfigure}[t]{.24\linewidth}
    \includegraphics[width=\linewidth]{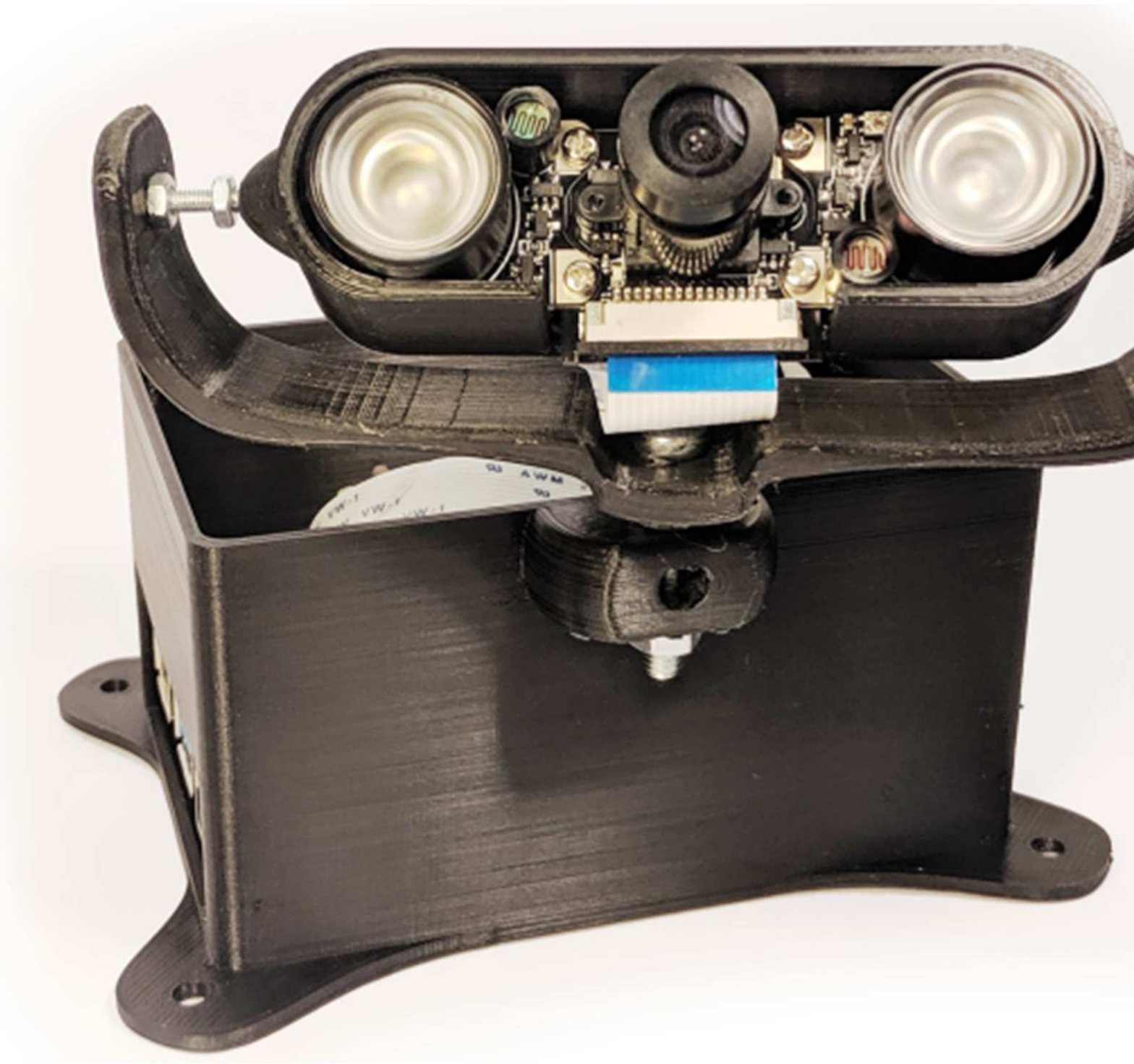}
    \caption{\label{f:deepdish}Intelligent}
  \end{subfigure}
  \caption{\label{f:sensorpics}Four types of sensors used in our deployment.}
\end{figure}

The three types of dumb sensor we deployed are depicted in Figure~\ref{f:sensorpics}, and are:

{\bf Smart plugs} (Figure~\ref{f:smartplug}). These are COTS smart plugs from several vendors built around the ESP8266 part~\cite{esp8266}. We replaced their default firmware with the Tasmota firmware~\cite{tasmota} so we could control where data were sent. The smart plugs were controlled over Wi-Fi using Message Queuing Telemetry Transport (MQTT).

{\bf LoRaWAN Sensors} (Figure~\ref{f:lorasensor}). We use different types of COTS LoRaWAN sensors, measuring e.g.,~CO$_2$, temperature, and occupancy from vendors including Elsys and Radio Bridge. These were managed over LoRaWAN via The Things Network (TTN).

{\bf ZigBee Sensors} (Figure~\ref{f:zigbeesensor}). We use two types of ZigBee sensor: infra-red motion sensors, and door/window open/closed sensors. These sensors were accessed via ZigBee gateways.

\paragraph{Intelligent Sensors}

The DeepDish~\cite{danish2020deepdish} (Figure~\ref{f:deepdish}) intelligent sensor counts the number of people in an area. DeepDish uses TensorFlow to identify and track selected objects (e.g.,~cars, bicycles, people) in the video feed, and supports occupancy counting by counting how many people cross a line in the scene in each direction. Only the cumulative occupancy count is reported, over Wi-Fi; no video data is collected or transmitted.

\subsection{Network Layer}

\begin{figure}
  \centering
  \begin{subfigure}[t]{\linewidth}
    \includegraphics[width=.5\linewidth]{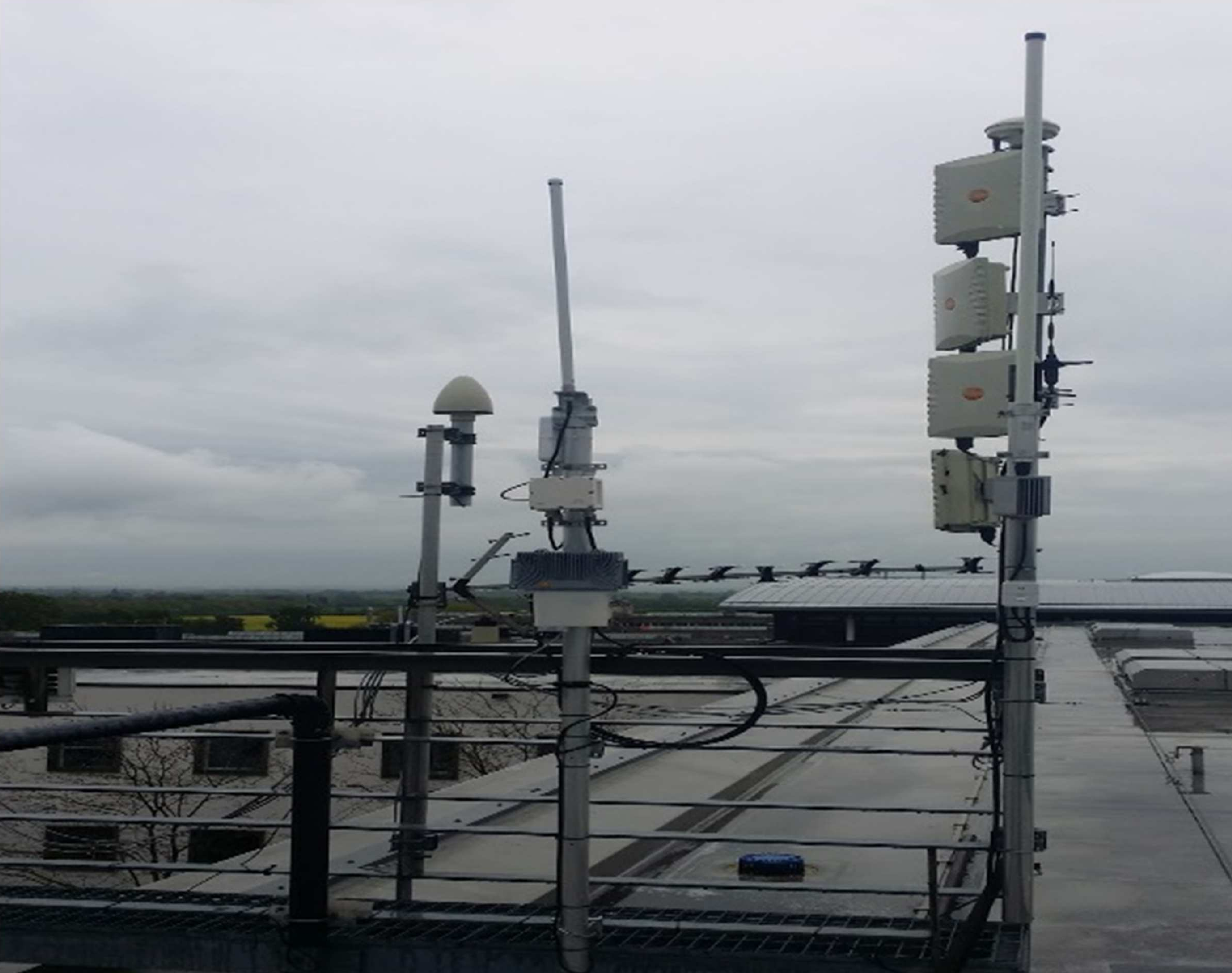}
    ~
    \includegraphics[width=.5\linewidth]{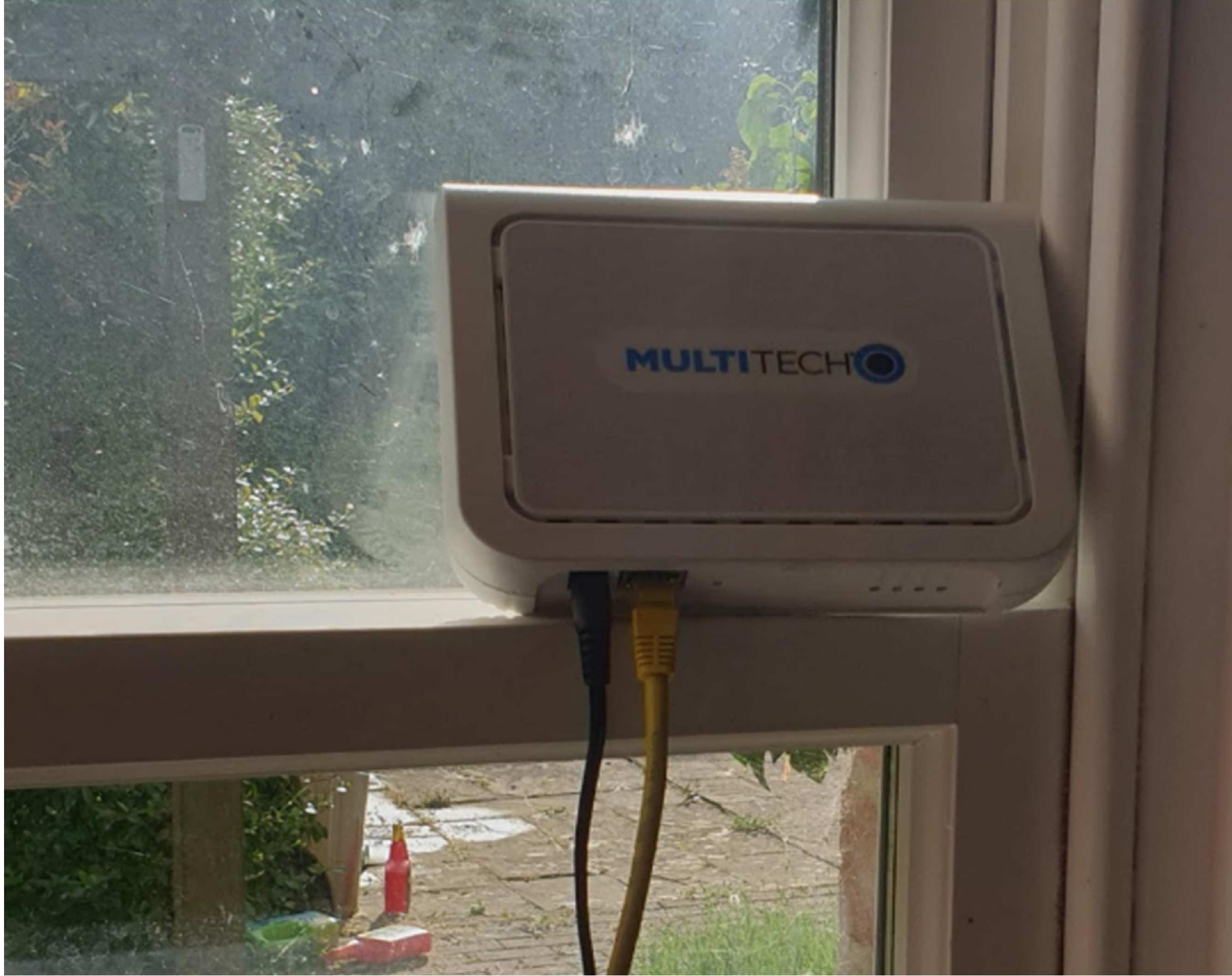}
    \caption{\label{f:lorawangateway}LoRaWAN gateways, on roof and in building}
  \end{subfigure}
  \begin{subfigure}[t]{.6\linewidth}
    \includegraphics[width=.5\linewidth]{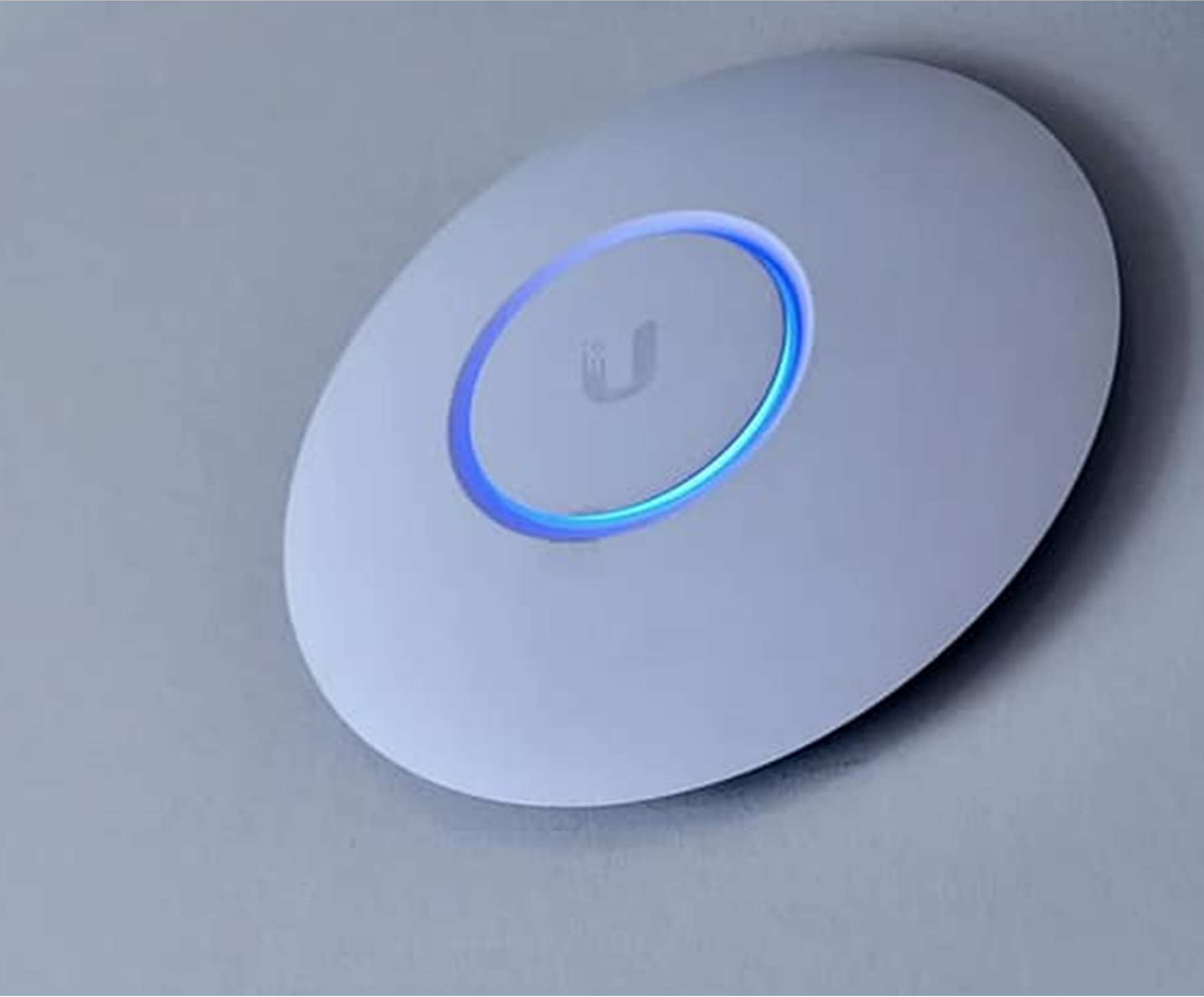}
    ~
    \includegraphics[width=.5\linewidth]{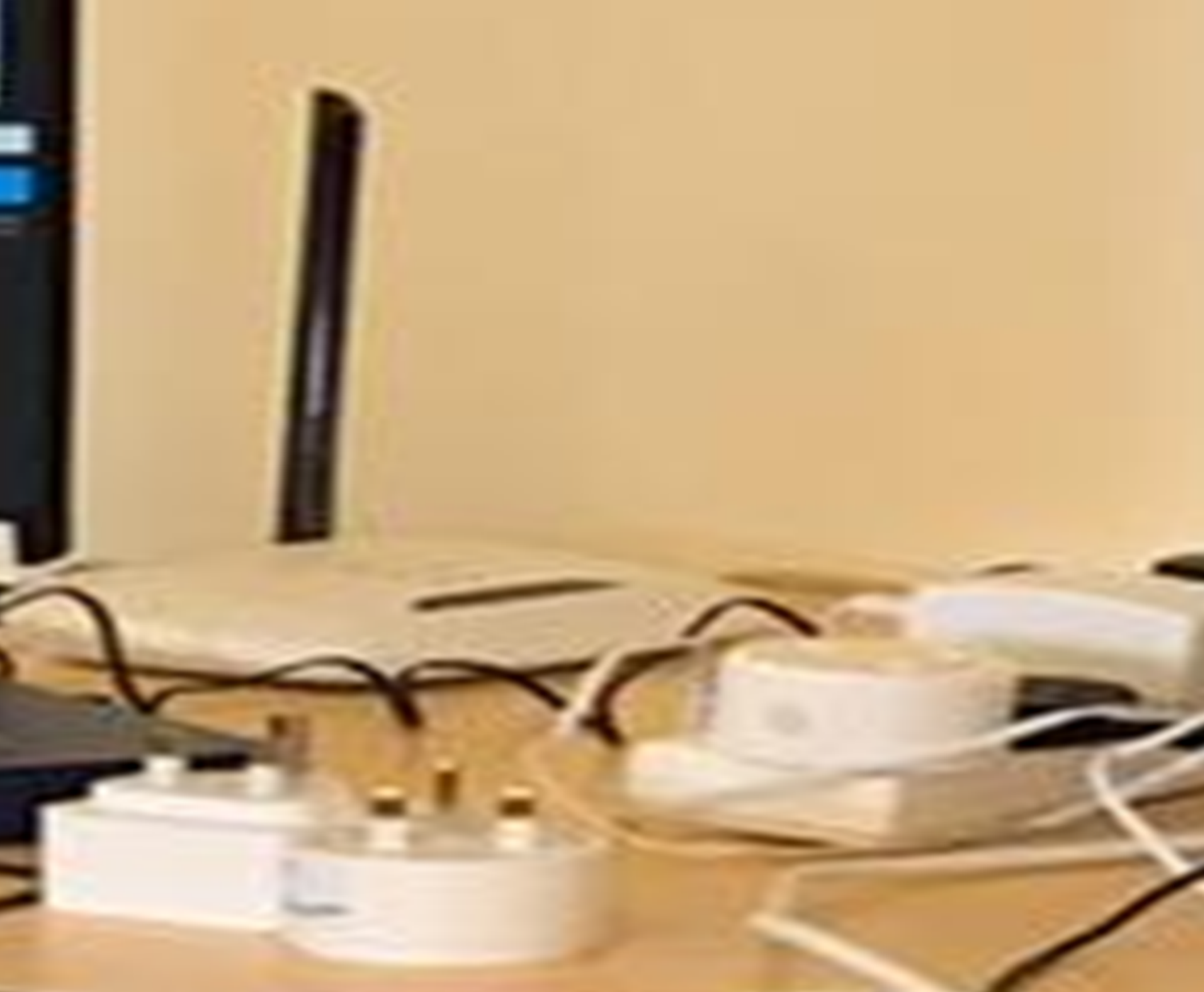}
    \caption{\label{f:wifigateway}Wi-Fi access points}
  \end{subfigure}
  \hfill
  \begin{subfigure}[t]{.3\linewidth}
    \includegraphics[width=\linewidth]{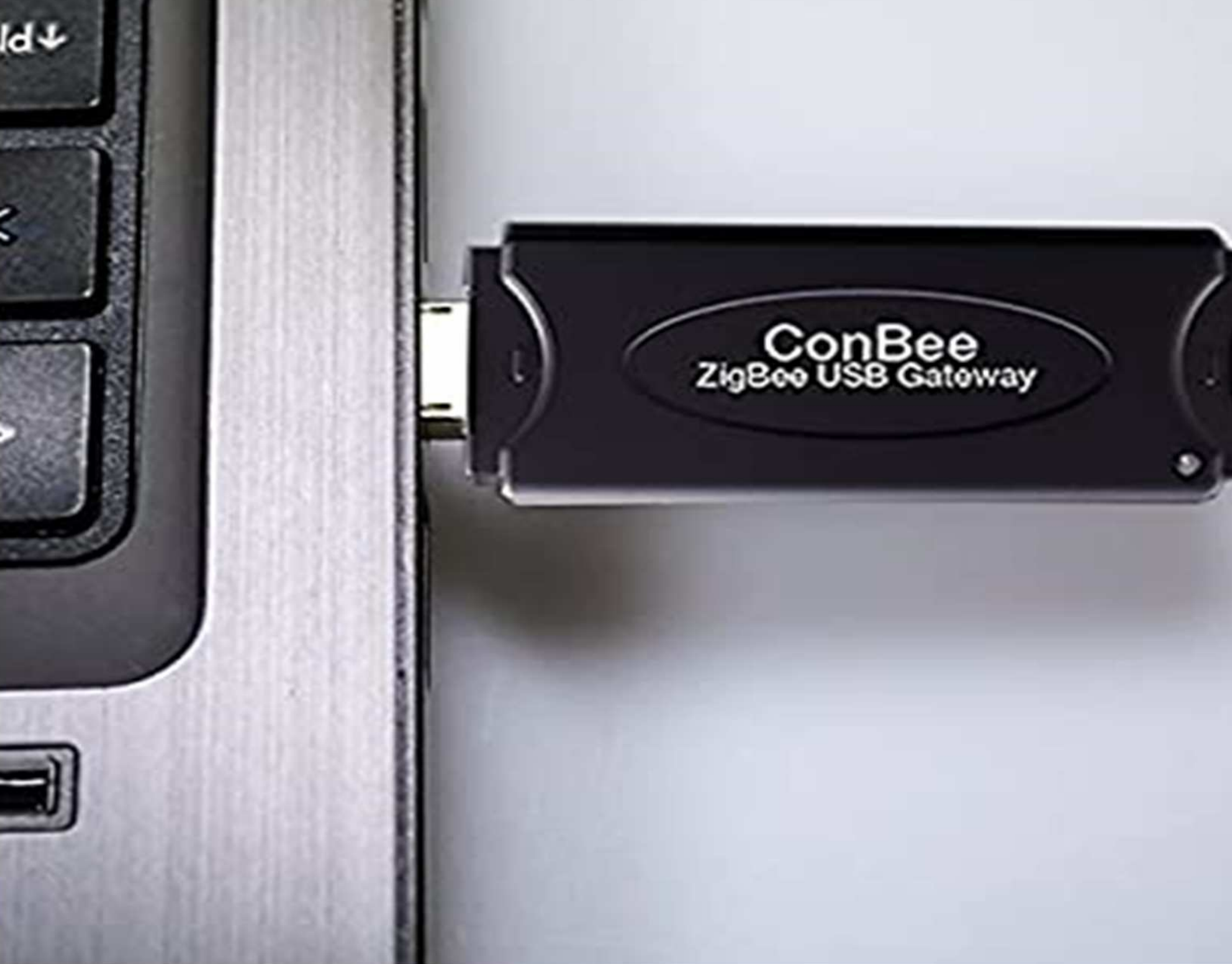}
    \caption{\label{f:zigbeegateway}ZigBee gateway}
  \end{subfigure}
  \caption{\label{f:netdevices}Deployed network devices.}
\end{figure}

The sensors used in our deployment required LoRaWAN, Wi-Fi, and ZigBee for data transfer.

{\bf LoRaWAN}. We deployed LoRaWAN gateways from  Semtech and Multitech on the rooftops and inside the buildings (Figure~\ref{f:lorawangateway}). Network access was made available via The Things Network (TTN). The hardware not only provided  long-range connectivity but also low power (sensors target several years lifetime on a single battery) and low bandwidth (51 bytes/message) support. The gateway used Wi-Fi to connect to TTN.

In our implementation the gateways were within 2 km range of all the sensors allowing for any transmission to use the \textit{spreading factor} suggested for real-time monitoring systems, $SF7$~\cite{adelantado2017understanding}. Such a low spreading factor guarantees low latency and supports around 1500 devices transmitting through the gateway on the same channel with a low packet error rate~\cite{lavric2019lora}.

{\bf Wi-Fi}. We used two classes of Wi-Fi APs. The first, from tp-link and D-link, follows the IEEE 802.11b/g/n standards and supports a maximum of 32 clients at a time and a maximum rate of 300 Mbps. The second, from Ubiquiti, follows the IEEE 802.11ac standard, supports a maximum of 250 clients, and provides a maximum rate of 450Mbps over the 2.4GHz frequency and 867Mbps over the 5 GHz frequency. In a practical setting based on factors like size, cost, range, one or both APs could be used. Based on the requirement, several of these APs were set up in different parts of the building  (Figure~\ref{f:wifigateway}).

\begin{figure}
  \centering
  \includegraphics[width=\linewidth]{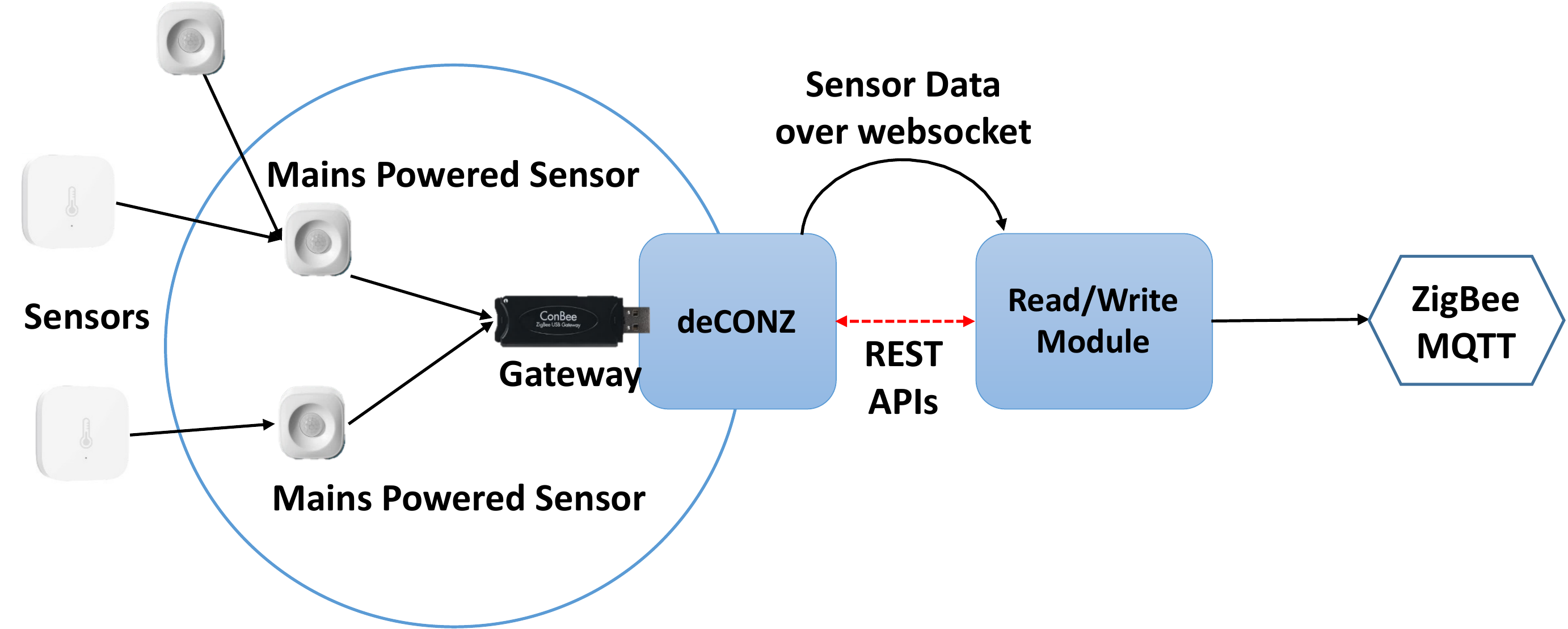}
  \caption{\label{f:zigbeemsg}ZigBee Sensor support framework.}
\end{figure}

{\bf ZigBee}. We used USB-based ConBee II gateways (Figure~\ref{f:zigbeegateway}) which do not require Internet access and have a range of up to 30 m inside. For more distant devices we created a ZigBee mesh network (Figure~\ref{f:zigbeemsg}) for which any ZigBee device connected to mains power acts as a repeater and routes signals.

\subsection{Integration Layer}

We used the Things Network (TTN) services, deCONZ~\cite{deconz}, and MQTT to integrate readings from the sensors. MQTT also served to bridge messages received over multiple channels, implementing the publish/subscribe design principle. We next describe the three phases of integrating the physical layer with the layers above.

\subsubsection{Smart plug Message Exchange}
We used the MQTT protocol for message exchange with the smart plugs. MQTT is a popular lightweight protocol for IoT projects, providing publish/subscribe support for real-time data exchange. MQTT has three primary components; \one~\emph{broker}: which is the server handling data exchange, \two~\emph{publisher}: a client which sends a message, and \three~\emph{subscriber}: a client which retrieves messages. Publishing and subscribing is with reference to a unique \emph{topic}, and a client can act both as a publisher or subscriber. We use the \emph{mosquitto} broker~\cite{eclipsemqtt}. Each smart plug publishes and subscribes to a topic unique to the device id. The broker receives periodic messages from all the sensors.

\subsubsection{ZigBee Sensor Message Exchange}
We set up our framework to obtain ZigBee sensor data through MQTT, as shown in Figure~\ref{f:zigbeemsg}. We used deCONZ~\cite{deconz}, an application which communicates with the gateway to expose the devices connected to the gateway. It provides websocket support for real-time data exchange and a set of APIs to manage sensors. We introduced the \emph{Read/Write} module, which continuously listens for data from the deCONZ websocket and publishes the same on the ZigBee MQTT broker.

\subsubsection{LoRa Sensor Message Exchange}
On deployment, The LoRa sensors are registered with the TTN using Over The Air Activation (OTAA). Once deployed, these sensors start broadcasting the LoRaWAN messages over the LoRa radio protocol, which is received by the deployed gateways. These gateways forward the LoRaWAN messages to TTN over the Internet, which is made available through the MQTT API. The messages obtained through the MQTT API are linked to a topic to which the sensors and the server subscribe. TTN also ensures that message transfer is secured.

\subsubsection{Bridging}
In our implementation, we had three channels, each having its own MQTT brokers. In order to integrate messages from all channels at a single broker, we utilise the MQTT bridging technique. Effectively, we subscribe the Wi-Fi MQTT broker (say \emph{localMQTT})  to the TTN MQTT broker (say \emph{ttnMQTT}) as well as the ZigBee MQTT broker (say \emph{zigbeeMQTT}) by  configuring \emph{localMQTT} appropriately. This configuration is flexible because we could decide if the bridging is required on all the topics or a subset of topics. Once set up, \emph{localMQTT} acts as the sole broker for data exchange between the Physical and other layers. This ensures that the Data layer need not tackle any changes in inclusion or removal of brokers in the architecture.

\subsection{Data Layer}
This handles three major tasks, \one~decoding and homogenising the data obtained from different channels via the \emph{localMQTT} broker, \two~storing data for future use, with associated metadata, and \three~making the data available in real-time for processing by client-side applications.

\subsubsection{Decoders}
The messages transmitted by the sensors vary based on factors like the type of sensor, vendor, design, transmission channel. For example, the smart plugs we used include their unique device number in the MQTT topic and not in the message, while most of the LoRa sensors include this information in the message itself. Moreover, although \our strives that the time of a reading is generated as far upstream the architecture as possible, preferably at the sensor, it is not always achievable. Several real-world sensors do not include time in the message as they do not have a real-time clock. Simply receiving and storing the message without assigning any time to it would make any further processing a complex task, especially for a real-time system. The decoders take care of these problems and generate a normalised message for each sensor. We implemented a set of decoders for the different classes of messages received from the \emph{localMQTT} broker. The decoder program consists of two main components, \one~the decoder set, which includes all the decoders for the available sensors, and \two~the decoder manager, which based on the message decides which decoder to use as well as automatically registering new decoders added to the decoder set. For our prototype, we simply append the timestamp and unique device id to all messages. However, more complex processing could be done to generate a more advanced payload.

\subsubsection{Storage}
\our has two types of storage: a data store containing all the data received from the physical devices in separate JSON files, and a metadata store containing metadata for all devices except network devices.

\begin{figure}
  \centering
  \includegraphics[width=\linewidth]{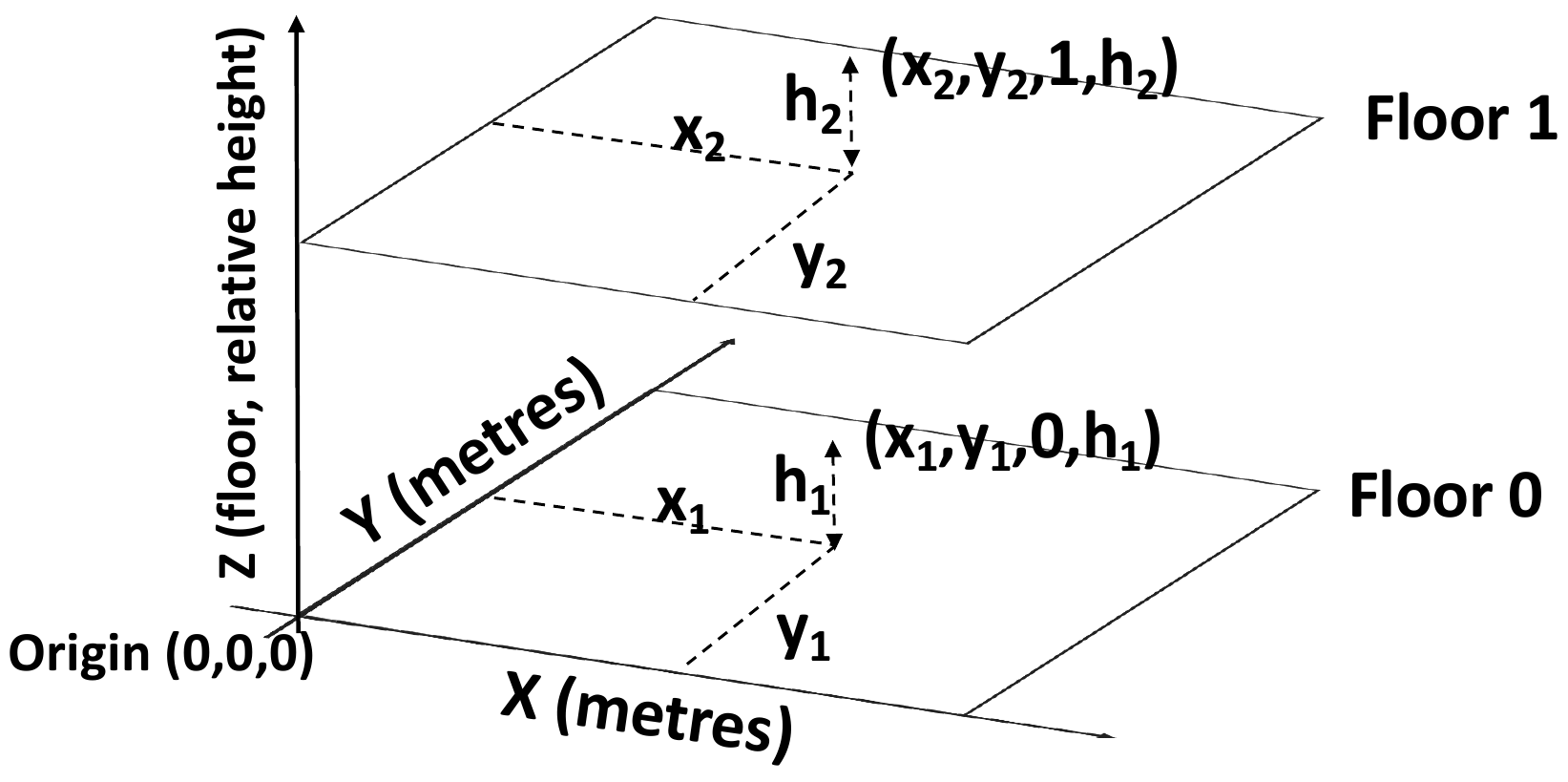}
  \caption{\label{f:coord}The in-building coordinate system followed in the implementation. X and Y coordinates are distance from origin in metres and Z coordinate is the combination of floor number and relative height on the floor.}
\end{figure}

We used PostgreSQL to implement the metadata store. As different devices have different metadata available, we use a simple two column table, one with the sensor's unique id, and the other a \emph{jsonb} column storing all other information in JSON format. The JSON entry for each device contains at least  a timestamp property, guaranteeing that we have all the historical changes of a sensor in the database instead of only having the latest information; and a location property, indicating where in the building the sensor is deployed. The location information also included the coordinates in the XYZ plane, with one corner of the building as the origin to calculate the XY coordinates, and the Z coordinate combines the floor the sensor is on and the height relative to that floor (Figure~\ref{f:coord}).

\subsubsection{Real-Time Server}
The Real-Time Server (RTS) supports minimal latency processing of real-time data and asynchronous \& non-blocking data management. This is guaranteed by following the \emph{Actor Model}~\cite{hewitt2010actor} using Vert.x~\cite{clement2015vert} to receive and support analysis of data in real-time by multiple Vert.x modules called \emph{verticles}. The \emph{actor model} supports concurrency by enforcing that each \emph{actor} (here implemented as a  verticle) only interacts with other actors (verticles) through messages posted to its \emph{message-box}. This model provides the following advantages:

\textbf{Asynchronous message passing}. The actor model guarantees that the RTS adheres to an asynchronous message-passing paradigm, providing real-time data handling from whichever source it is received. The \emph{FeedHandler} receives a message arriving at the RTS and publishes on the \emph{EventBus} to be used by other verticles.

\textbf{Non-blocking modules}. Using the Vert.x library with the actor model provides support to build a non-blocking framework for the RTS. Unlike the standard actor model, where each actor has its own \emph{message-box}, in our implementation, the \emph{EventBus} acts as the common \emph{message-box} for all verticles. The message exchange is performed using a publish/subscribe approach. Any verticle accesses data by subscribing to the \emph{EventBus} and sends messages by publishing it to the \emph{EventBus}. The model also ensures that any communication between two verticles also happens only through the \emph{EventBus}.

\textbf{Modular server}. Each verticle is an independent actor ensuring that the RTS is modular. This guarantees that any number of modules could be added or removed as needed without affecting any existing verticles. As a result, a standard production implementation could have thousands of verticles accessing data concurrently.

As well as following the actor model for concurrency, all verticles in \our act as \emph{stream processors}. Unlike most systems, which store the data in a storage unit and then query or perform computation over it, stream processing differs in two key ways:

(1)~\textbf{Events substitute messages}. Verticles in \our react to the incoming stream of events instead of a message or a batch of messages. Many sensors will send periodic updates reporting the status quo -- these are typically not of interest to verticles, which are concerned rather with events indicating some change of state. For instance, a verticle controlling lights in a room might only be interested in the events indicating a change from unoccupied to occupied, or vice versa, and not in processing periodic messages of current occupancy which the sensor sends. Working with events also ensures that \our handles \emph{timeliness} of the event being processed by a verticle.

(2)~\textbf{Reversing the norm}. Unlike the store-then-analyse approach, stream processing focuses first on enabling real-time reactive processing of data (events). Upon receiving a relevant event, a stream processing application (a verticle in \our) reacts to the event by updating some information, creating another event, or simply storing it. The result is that data can still be archived for historical processing, but this does not negatively affect the performance of real-time processing.

There are four classes of verticles that \our requires: \one~\emph{Data ingestion verticles}, which receive data from the MQTT broker and publish the same on the \emph{EventBus} (e.g.,~\emph{FeedHandler}), \two~\emph{Data storage verticles}, which subscribe to the \emph{EventBus} for any new data and store it for future usage (e.g.,~\emph{MessageFiler}), \three~\emph{Real-time analysis verticles}, which analyse the stream of data in real-time and publish updates on the \emph{EventBus}, and \four~\emph{Outbound verticles} which make the data available to the outside world. As shown in Figure~\ref{f:implementation}, additional verticles include the \emph{MessageRouter} verticle, used to share data to other similar systems; the \emph{Data Monitor}, used to interact with client-side applications; and the \emph {Real-Time Analysis} verticle, comprising one or more verticles and performing tasks like identifying events such as the measured CO$_2$ level crossing a threshold or a power outages, and broadcast the results of such analysis as derived events on the \emph{EventBus}.

We set up three backup servers, which could act as the primary server during any outage. These backup servers subscribed to the \emph{EventBus} to receive any new message from a sensor. This ensured that the backup server had all the data that the primary server had.

\section{Case Study}\label{s:casestudy}

We next examine an end-to-end example of a system using \our to fuse sensor data to provide a useful application: a modernisation of the \emph{Trojan Room Coffee Pot}~\cite{coffee}. The original deployed one of the first webcams to monitor how full was a research group's coffee pot. 

\begin{figure}
  \centering
  \begin{subfigure}[t]{\linewidth}
    \includegraphics[width=.95\linewidth]{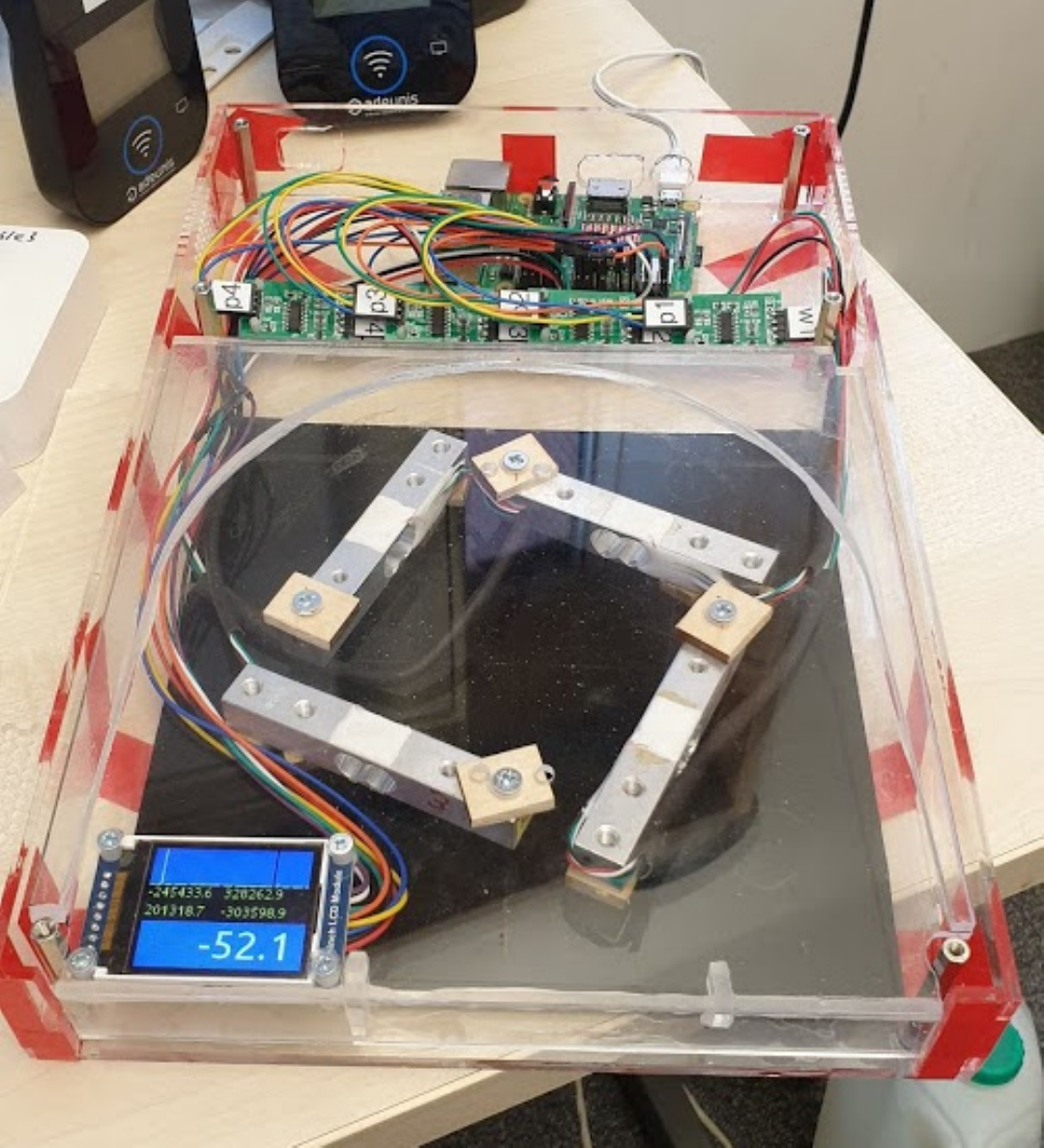}
    \caption{\label{f:coffeenode}Coffee pot sensor node comprising four weight sensors, a Raspberry Pi and Wi-Fi gateway, with a local display for convenience.}
  \end{subfigure}

  \begin{subfigure}[t]{\linewidth}
    \includegraphics[width=\linewidth]{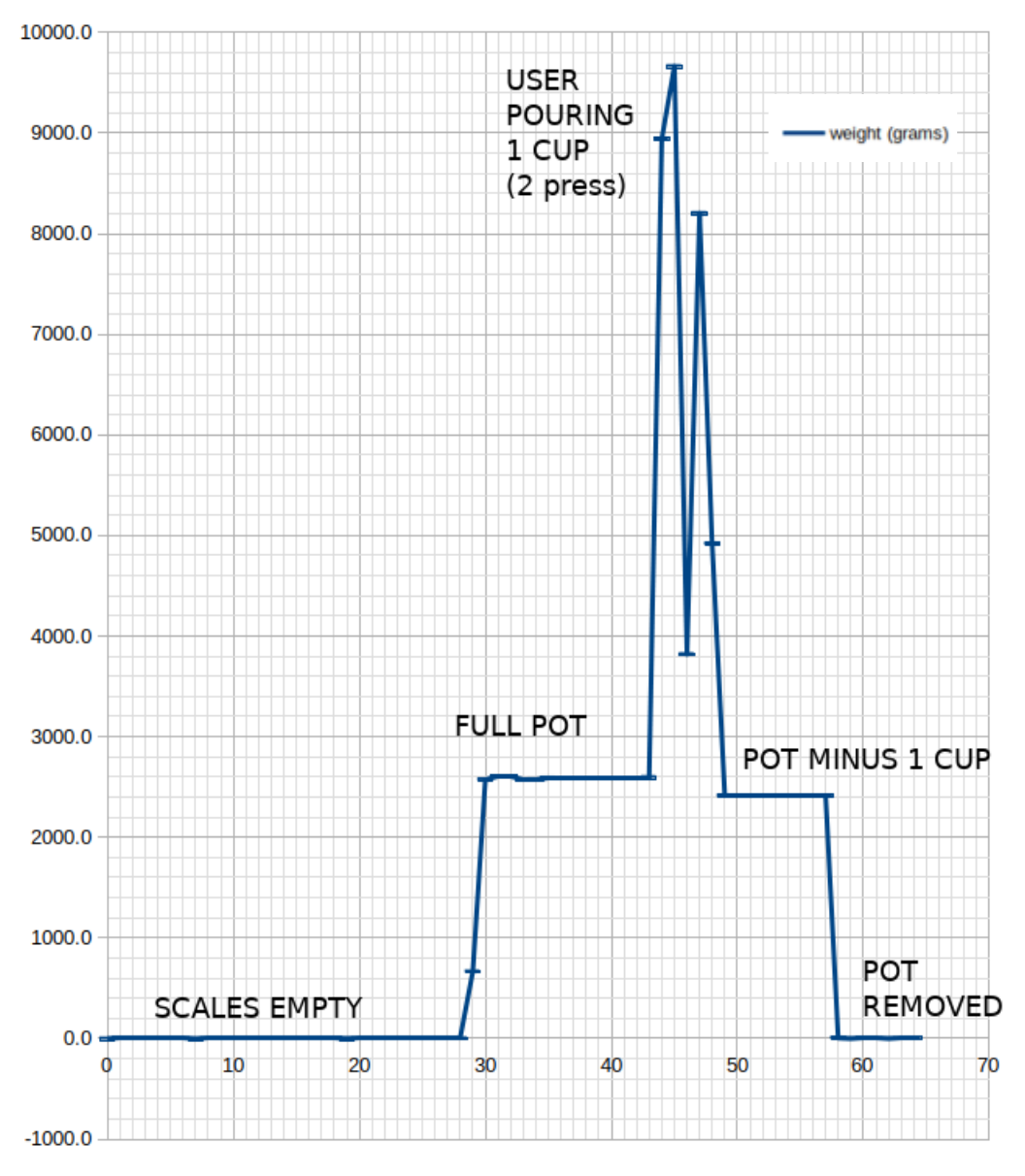}
    \caption{\label{f:coffeereadings}Readings during coffee preparation.}
  \end{subfigure}

  \caption{\label{f:coffeepot}The Coffee Pot sensor node. The pot weighs $\sim$0.5 kg, and when full holds 2 kg of coffee. Each cup taken is $\sim$0.25 kg.}
\end{figure}

In our modernised version, where an opaque coffee pot renders the webcam approach ineffective, we measure and transmit in real-time the coffee-making and consuming events of the coffee pot in one of our buildings. The system analyses data from a set of sensors deployed at the coffee pot to recognise one of five events; \one~{\tt pot-removed}, indicated the pot is not present, \two~{\tt new-pot}, indicating the presence of freshly-made coffee in the pot, \three~{\tt pot-poured}, indicating that coffee has been poured, \four~{\tt pot-empty}, indicating that no coffee remains, and \five~{\tt coffee-grinding}, indicating that the coffee bean grinding machine appears to be active.

The coffee pot setup (Figure~\ref{f:coffeenode}) is designed as a \emph{sensor node}, a coordinated collection of multiple sensors: weight sensors connected to the Raspberry Pi periodically monitor the weight of the coffee pot, while two smart plugs monitor the power usage of the grinder and the coffee brewing machine respectively. The Raspberry Pi also provides a Wi-Fi gateway to connect the two smart plugs. The sensor node accumulates data from each sensor and transmits a message to the local MQTT broker over Wi-Fi. The message consists of the weight and power readings and the time when the reading was recorded.

\begin{figure}
  \centering
  \includegraphics[width=\linewidth]{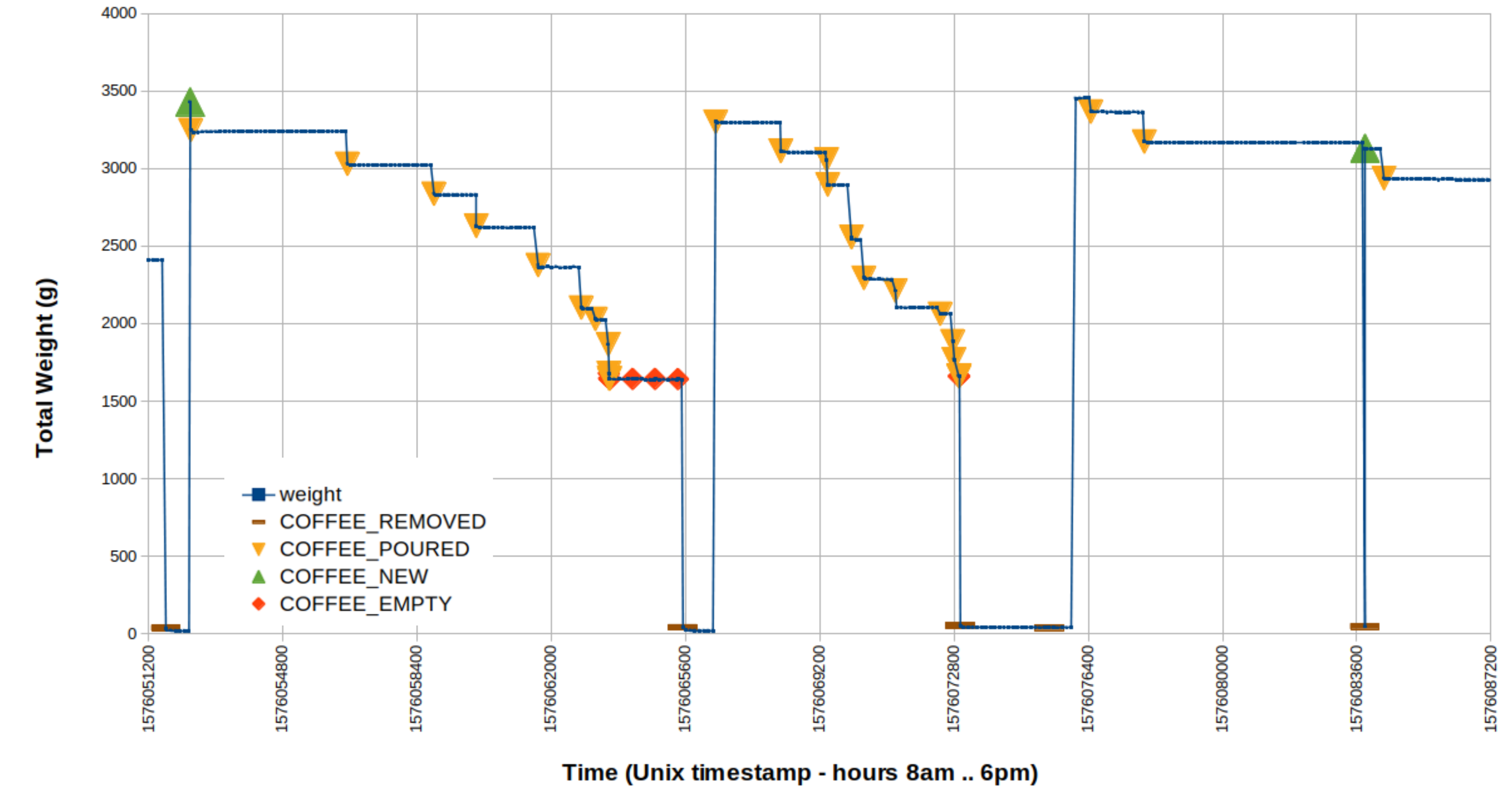}
  \caption{\label{f:rtevents}Events observed by \emph{RTCoffee}.}
\end{figure}

After being homogenised by the corresponding decoder, the message is published on the \emph{EventBus} by the \emph{FeedHandler} verticle. The \emph{MessageFiler}, subscribed to the \emph{EventBus} for any new message receives the new message and stores the attached data. A real-time analysis verticle, \emph{RTCoffee}, looks for two types of events in the published data: did the power consumed by either the grinder or the coffee machine cross a threshold (40\,W in our case), indicating the grinder or the coffee machine was in use; and has the measured weight of the coffee pot changed, indicating one of the five events described above, as depicted in Figure~\ref{f:coffeereadings}. Figure~\ref{f:rtevents} gives an example of the events observed by \emph{RTCoffee} on a particular day between 8am and 6pm. \emph{RTCoffee} publishes its derived event on the \emph{EventBus} to be consumed by other verticles.

\begin{figure}
  \centering
  \begin{subfigure}[t]{.32\linewidth}
    \includegraphics[width=\linewidth]{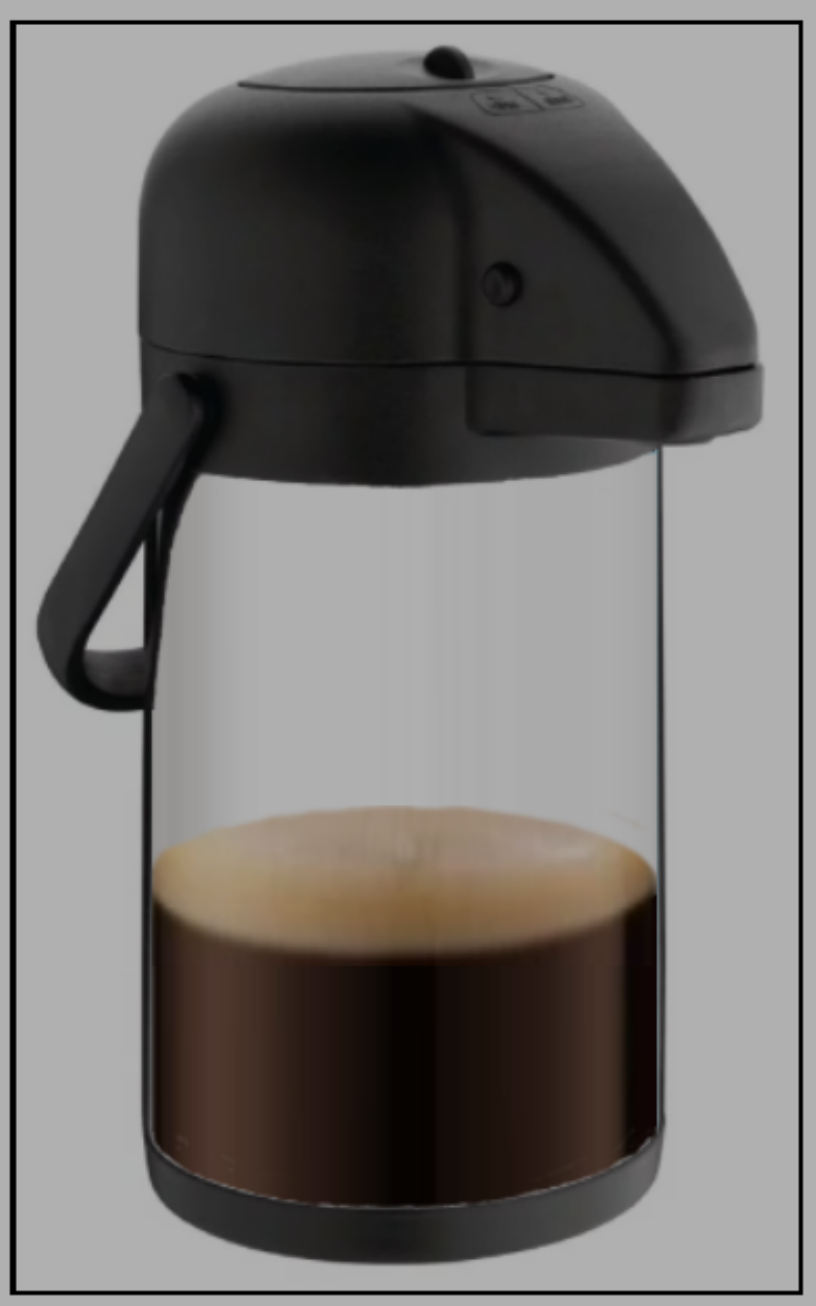}
    \caption{\label{f:coffeelevel}The level of coffee in the pot at a given time.}
  \end{subfigure}
  \hfill
  \begin{subfigure}[t]{.32\linewidth}
    \includegraphics[width=\linewidth]{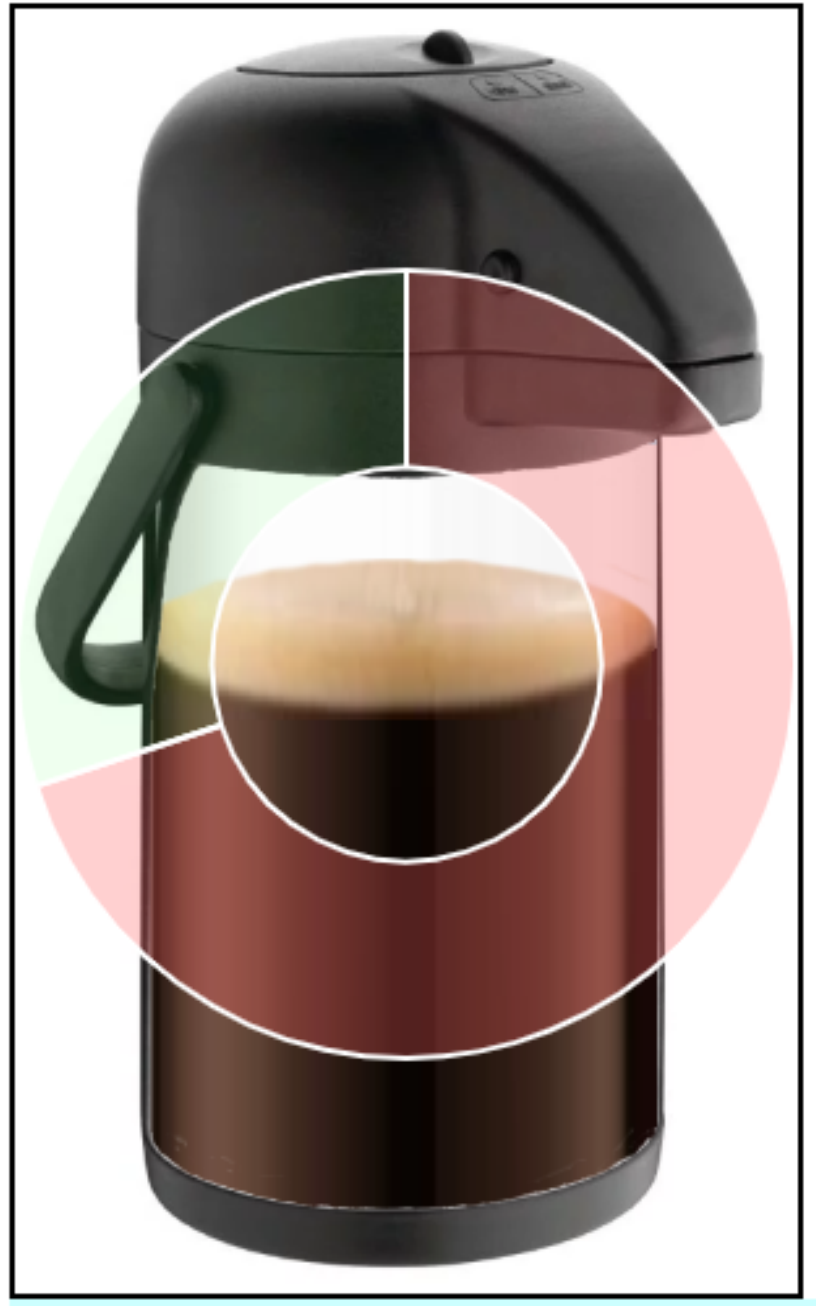}
    \caption{\label{f:coffeestatus}Level of coffee in the pot after a
      \textit{coffee-grinding} event with approximate time remaining.}
  \end{subfigure}
  \hfill
  \begin{subfigure}[t]{.32\linewidth}
    \includegraphics[width=\linewidth]{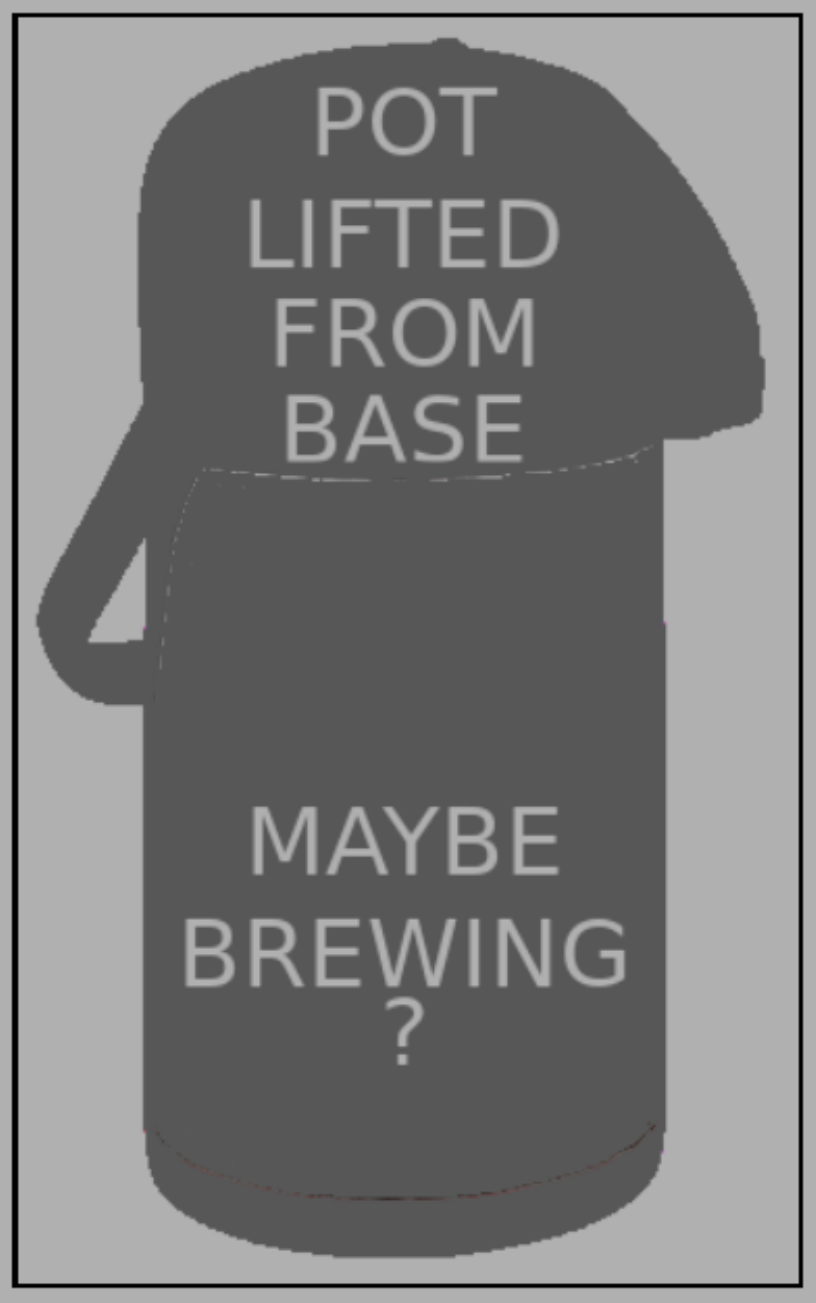}
    \caption{\label{f:coffeeremoved}Following a \textit{pot-removed} event.}
  \end{subfigure}
  \caption{\label{f:coffeepotui}Web-Client UI showing status of the coffee pot.}
\end{figure}

We implemented a simple web client to consume the data received from \our and display the status of coffee in the coffee pot. The web client subscribes to a \emph{DataMonitor} verticle for event updates, which in turn subscribes to the \emph{EventBus} and receives any updates provided by \emph{RTCoffee}. Whenever an update occurs, the web client receives the event, and the UI is updated accordingly. Some example UI updates are shown in Figure~\ref{f:coffeepotui}.

\section{Evaluation}\label{s:evaluation}

Our deployment of \our has been live since March 2020, and we present measurements taken over seven months, to October 2020. We show that \our provides an architecture for minimal latency real-time data processing. Many sensors were added during the experiment period, and we examine how this impacted performance. We then compare against some of the existing alternative systems to highlight the advantage of the design choices made in \our.

\subsection{Real-time Performance}
As a real-time architecture, it is important for \our to show minimal latency between the generation of data at the sensor and the consumption of data by an application. In order to validate this, we measured the latency at four key points in the \our architecture: \one~the gateway receiving the first hop message from a sensor \two~the bridge MQTT broker in the Integration layer, \three~the EventBus in the Data layer, and \four~a client-side application similar to that described in~\s{s:casestudy}.

\begin{table}[!ht]
  \begin{tabular}{c|c|c|p{1cm}|c|}
    & \textbf{Gateway} & \textbf{\begin{tabular}[c]{@{}c@{}}MQTT \\ Broker\end{tabular}} & \textbf{Event- Bus} & \textbf{\begin{tabular}[c]{@{}c@{}}Client\\ Application\end{tabular}} \\
    \hline
    \textbf{Mean} (ms) & 57.15   & 147.86      & 157.86   & 159.55 \\
    \hline
    \textbf{Std Dev} (ms) & 10.21   & 63.56       & 2.35     & 0.56 \\
    \hline
  \end{tabular}
  \caption{Latency at key points of data flow in the \our architecture. All values are in ms and are calculated from the time of message generation at the sensor.}
  \label{tab:latencyall}
\end{table}

As is evident from Table~\ref{tab:latencyall}, at all four points latency is minimal. With as low as $57 ms$ average latency at the gateway, an increase is observed at the integration layer (approx $90 ms$ in average) owing to the processing involved with the bridging of information from different protocols. The messages are published at the \textit{EventBus} with a latency of just $10 ms$, while the client-side application receives the desired data with almost no latency ($2 ms$ in average). For all practical purposes, an average latency of $159.55 ms$ is a negligible value which validates \our as a real-time system.

\begin{figure}
  \centering
  \begin{subfigure}[t]{\linewidth}
    \includegraphics[width=\linewidth]{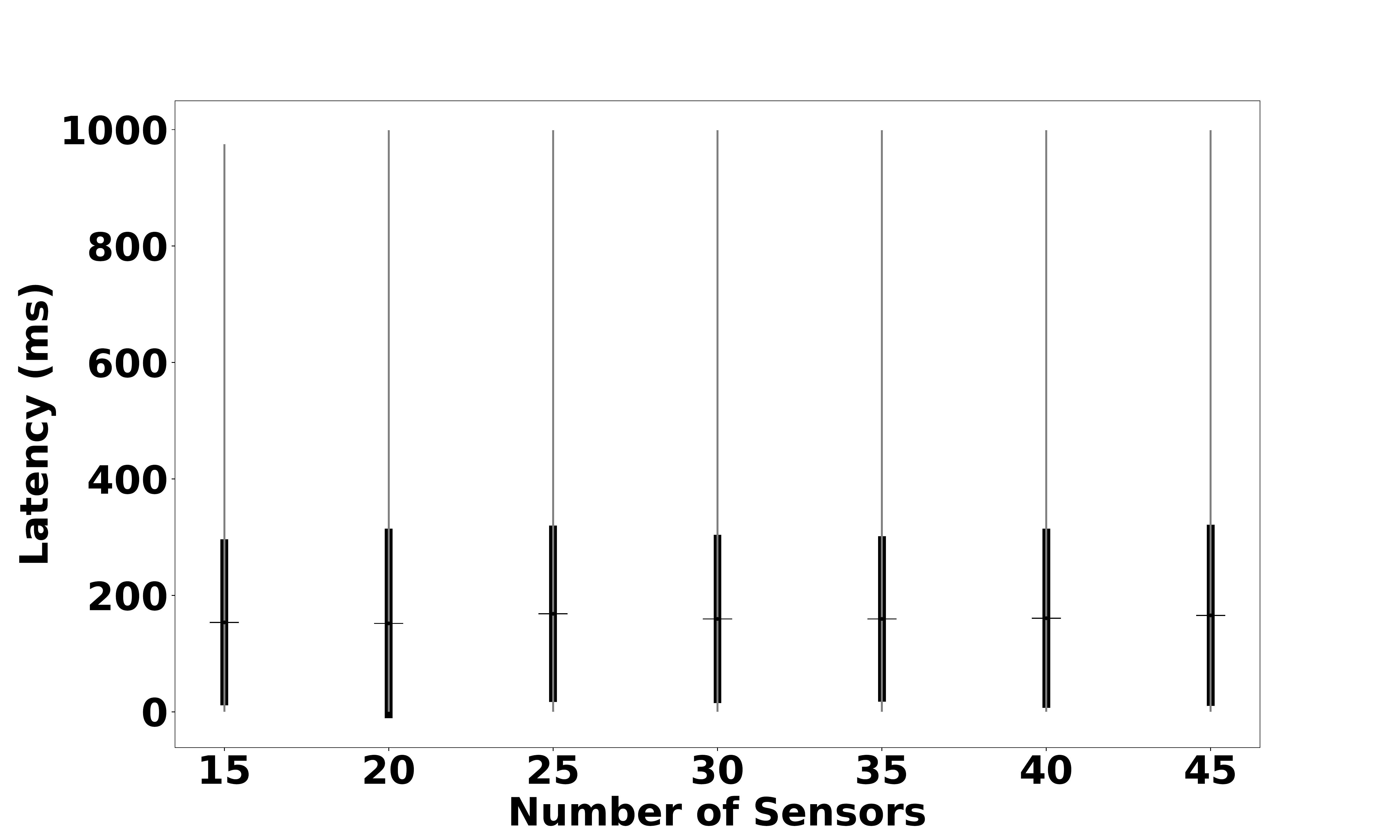}
    \caption{\label{f:latencysensors}Latency variation with sensor count}
  \end{subfigure}

  \begin{subfigure}[t]{\linewidth}
    \includegraphics[width=\linewidth]{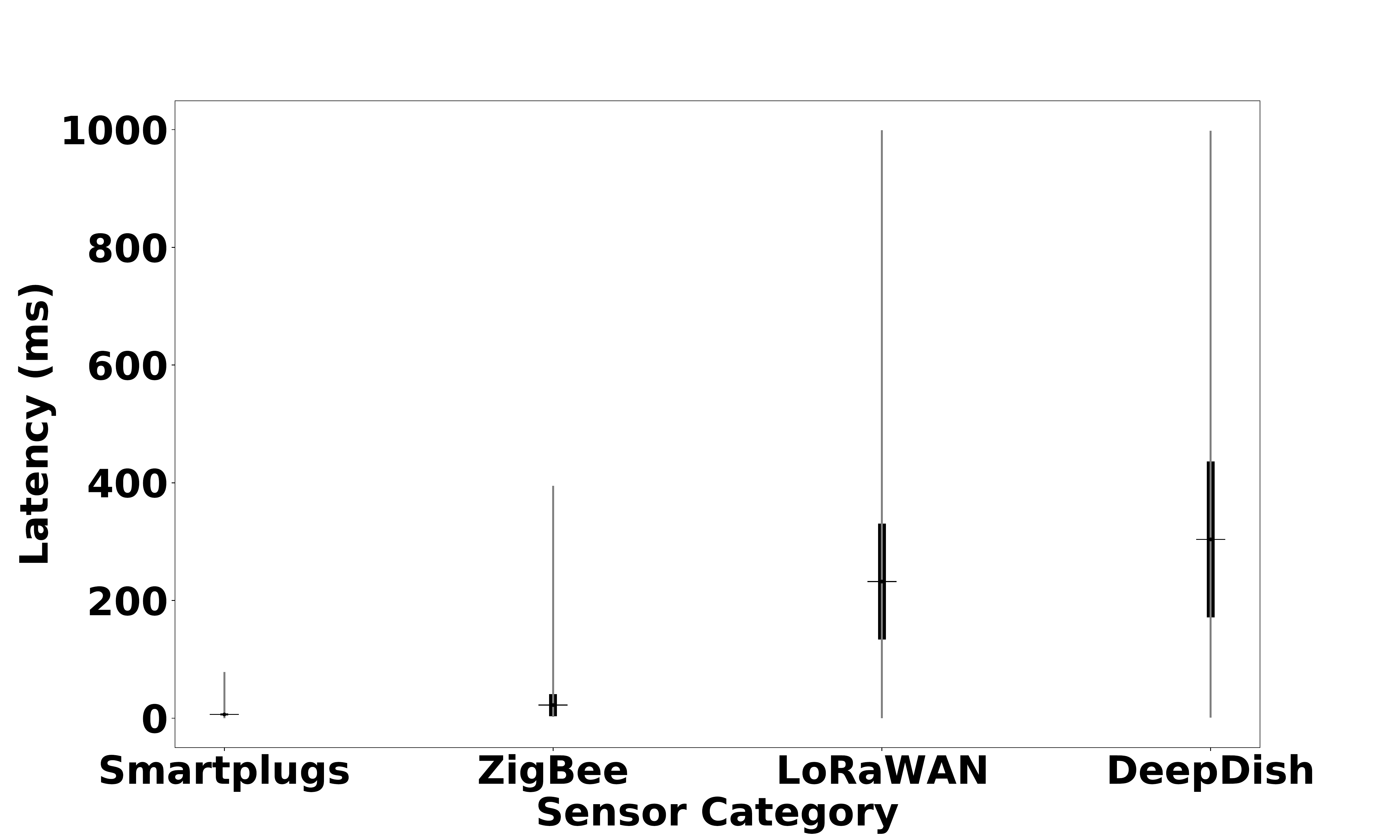}
    \caption{\label{f:latencycategory}Latency variation with sensor category}
  \end{subfigure}
  \caption{\label{f:latencyarch}Observed latency variation.}
\end{figure}

Furthermore, \our is robust enough that addition of new sensors or category of sensor in the system doesn't affect the overall working of the architecture. It is clear from Figure~\ref{f:latencysensors} that even on increasing the number of sensors in the system, the overall latency remains close to a similar mean value of $~200 ms$. This is true also for when we calculated the mean latency for the different category of sensors used in the implementation of \our (Figure~\ref{f:latencycategory}). Here, the maximum average latency is observed for \textit{DeepDish} ($400 ms$) which is because of $~200 ms$ processing time involved in computing the number of people from a video frame.

\subsection{Comparison Against Alternatives}
We compare different aspects of \our with three similar systems. The first is the work by Al-Ali et al.~\cite{al2017smart} which sets up a Wireless Sensor Network with each sensor interfaced with a data acquisition system on a chip. The sensors exchange data through a MQTT broker which is then sent to a central server for analysis. Any analysis in this system is request/response based. The second system by Fayyaz et al.~\cite{fayyaz2019iot} does something similar, however uses a fog server as the first hop point for the sensor messages. The third work we compare with is developed by Evangelatos et al~\cite{evangelatos2012evaluating} which uses IPv6 over Low power Wireless Personal Area Networks (6LoWPAN) to collect sensor information and store it for further processing. However, data is only requested for events such as a person entering a room, so no real-time data flow is required.

We compare systems on two fronts, \one~latency at the first hop point for the sensor messages (at the gateway), and \two~latency for a complete data flow transaction from sensor to application. Both results were not reported for all the three competing systems, hence we only include the ones which were reported for each criteria. Also, the results for Fayyaz et. al.~\cite{fayyaz2019iot} are based on simulation.

\begin{table}
  \begin{tabular}{l|c|c|c|c}
    {\textbf{System}} & \textbf{\our} & \textbf{~\cite{al2017smart}} & \textbf{~\cite{fayyaz2019iot}} & \textbf{~\cite{evangelatos2012evaluating}} \\ \hline
    \textbf{Sensors Used}                 & 45               & 10            & 12           & 20          \\ \hline
    \textbf{First hop Latency (ms)}       & 57               & 66            & 60           & -           \\ \hline
    \textbf{End-to-End Latency (ms)}      & 160              & -             & 6000         & 259         \\ \hline
  \end{tabular}
  \caption{\label{tab:competing}Comparing state-of-the-art systems with \our.}
\end{table}

\paragraph{First hop latency} We calculated the average first hop latency observed over all category of sensors which we report in Table~\ref{tab:competing}. It is evident that the latency is at par with the state-of-the-art values (and in fact it beats them by a few ms).

\paragraph{End-to-End latency} When compared to the fog-based system~\cite{fayyaz2019iot}, \our performs 37 times better. This is primarily because the application server subscribing to the Data Monitor and hence receiving the messages almost at the same time as the EventBus (2 ms latency, as shown in Table~\ref{tab:latencyall}). However, that system stores all the data on the cloud and then queries that from every application, increasing the latency. The 6LoWPAN system~\cite{evangelatos2012evaluating} does achieve almost similar latency as \our, but only for the data stored at the server. Additional analysis would have further increased latency.

\section{Discussion}\label{s:discussion}

\our provides a robust architecture for in-building real-time data flow but there remain some key elements requiring further work.

\subsection{Scalability}
This is more a feature of our prototype implementation than the \our architecture itself. The prototype implemented in this paper covers only two buildings so far, with two more buildings in the deployment pipeline. It  will be essential to observe the impact on performance as this number is increased.

As well as extending to cover more buildings, we are extending the types of sensors used in the implementation. We would like to integrate other types of data sources like Modbus or Monnit sensors~\cite{monnit}, and understand how to fit them within the architecture.

As the set of sensors being deployed increases, optimising the number of sensors in the building becomes important. For instance, if a ZigBee and a LoRaWAN sensor provide the same readings, only one might be used dependent on the client's needs. As sensors are increasingly integrated, a single sensor might provide multiple readings and so could replace multiple sensors. With more sensors in place, deciding the optimal placement strategy for effective building coverage is essential. Others have investigated this~\cite{lee2019optimal,papadopoulou2016optimal}, providing directions for finding a strategy for \our.

\subsection{Privacy}
Privacy is a key concern with IoT-based systems, and many have looked into how to achieve this in smart buildings and smart cities~\cite{martinez2013pursuit,pappachan2017towards,kim2017toolkit,jiang2019lightweight}. Currently, \our provides basic privacy support through encryption provided by network protocols and  limited data access. However, \our is amenable for more fine-grained privacy approaches. Our intended approach has three aspects.

First, we will ensure all messages from any sensor are encrypted. As some sensors do not have the capacity to encrypt data at source, decisions will need to be made about where is the most appropriate point in the architecture to provide encryption, and what scheme should be used.

Second, we need to define mechanisms to support different data access strategies governing who can access what data. For example, a person might have access to all the data generated by sensors in her room, but only specific sensors on the floor. This could involve assigning the different stakeholders, building managers, third-party clients, or building occupants, into groups by which access is controlled. Another option could be to provide unique tokens to each stakeholder, and access is provided based on tokens. This could also include building visitors to whom limited data could be made available, e.g.,~in a time of COVID-19 with social distancing recommendations, current (but not historical) room occupancy could help keep visitors and occupants safe.

Third, under whatever access control regime is provided, we must still determine how occupants' privacy should be protected. For instance, a building manager receiving power readings from smart plugs in a room every hour could infer when the occupant was in their office, whereas receiving overall power usage for an office over a week might be sufficient for managing energy efficiency in a building. We thus need to examine how client applications can receive sensor data so as  to ensure privacy while still meeting the differing goals of users of the system.

\section{Conclusion}\label{s:conclusion}

As the world moves towards more and more smart buildings, we anticipate considerable (perhaps exponential) increase in the number of sensors and consequently volumes of real-time data generated. Building Management Systems (BMSs) need to be re-architected to better support both data management and reliable real-time information dissemination.  \our provides a robust architecture satisfying these goals by considering several key aspects:  \one~spatio-temporal aspects of real-time sensor data, \two~improving \emph{timeliness} and maintaining low \emph{latency} throughout, \three~ensuring a homogeneous data flow through the architecture notwithstanding the wide range of sensor types and capabilities, and \four~performing efficient real-time analysis of sensor data with minimal latency.

Our implementation of a prototype and the experiments carried on for seven months show that \our does provide an efficient network architecture for in-building real-time data flow and real-time data analysis. There do exist key aspects of scaling and privacy, which would improve \our further. However, as it stands, \our is a first step towards providing a robust network architecture that could tackle the incoming challenges BMS faces regarding the increasing volume of sensors and the real-time data being generated by these sensors every day.


{
  \bibliographystyle{ACM-Reference-Format}
  \bibliography{refs}
}

\typeout{get arXiv to do 4 passes: Label(s) may have changed. Rerun}

\end{document}